\documentclass[12pt,preprint]{aastex}

\slugcomment{}
\def\J{J1420$-$0545 }
\def\D{{\sc DYNAGE} }
\def\A{{\sc AIPS} }
\def\St{{\sc STARLIGHT} }

\shorttitle{Understaning \J}
\shortauthors{Machalski et al.}

\begin{document}

\title{Understanding Giant Radio Galaxy J1420$-$0545:\\ Large-Scale Morphology, Environment, and Energetics}

\author{J. Machalski$^1$, M. Jamrozy$^1$, {\L}. Stawarz$^{2,\,1}$, and D. Kozie{\l}-Wierzbowska$^1$}

\affil{$^1$Astronomical Observatory, Jagellonian University, ul. Orla 171, 30-244 Krak\'ow, Poland}
\affil{$^2$Institute of Space and Astronautical Science, JAXA, 3-1-1 Yoshinodai, Sagamihara, Kanagawa, 229-8510, Japan} 

\email{machalsk@oa.uj.edu.pl, jamrozy@oa.uj.edu.pl, stawarz@astro.isas.jaxa.jp, eddie@oa.uj.edu.pl}

\begin{abstract}
In this paper we consider the possibility that the structure of the largest radio galaxy \J is formed by a restarted rather than a primary jet activity. This hypothesis is motivated by the unusual morphological properties of the source, suggesting almost ballistic propagation of powerful jets in a particularly low-density environment. New radio observations of \J confirm its morphology consisting of only two narrow lobes; no trace of any outer low-density cavity due to the previous jet activity is therefore detected. Different model fits performed using the newly accessed radio data imply relatively young age of the source, its exceptionally high expansion velocity, large jet kinetic power, and confirm particularly low-density environment. We find that it is possible to choose a realistic set of the model parameters for which the hypothetical outer lobes of \J are old enough so that their expected radio surface brightness is substantially below the rms noise level of the available radio maps. On the other hand, the extremely low density of the gas surrounding the \J lobes is consistent with the mean density of the baryonic matter in the Universe. This suggests that the source may be instead located in a real void of the galaxy and matter distribution. In both cases the giant radio lobes of \J are expected to modify substantially the surrounding matter by driving strong shocks and heating the gas located at the outskirts of the filamentary galactic distribution. Finally, we also find that the energetic requirements for the source are severe in terms of the total jet power and the total energy deposited by the outflows far away from the central engine.
\end{abstract}


\section{Introduction}

Giant radio galaxies (hereafter GRGs) are luminous but low-surface brightness sources dominated at radio wavelengths by the emission of extended lobes with linear sizes $\gtrsim 1$\,Mpc. Morphologically, most of GRGs resemble powerful Fanaroff-Riley type II objects \citep[FR\,IIs;][]{fan74}, and hence dynamical models developed for classical doubles are typically used to infer their intrinsic parameters based on the multifrequency radio data obtained using a variety of radio instruments \citep[e.g.,][]{kon08,jam08,mach09}. The key difficulty in modeling and understanding of such sources is however the fact that the main characteristics of the ambient medium surrounding giant lobes are hardly known. In fact, there are almost no observational constraints on the structure, temperature, or density of the intergalactic gas at Mpc distances from the centers of poor clusters or groups in which powerful radio galaxies, including GRGs, are expected to be located \citep{jam04,sar05}. But one may also turn the argument around, and argue that GRGs can indeed be used \emph{as unique probes} of the intergalactic medium (IGM) and of its cosmological evolution at the outskirts, or even outside of groups and clusters of galaxies. This idea was discussed and pushed forward by, e.g., \citet{sub08} and \citet{saf09}.

Clearly, using GRGs as cosmological probes requires a consensus regarding the nature of those objects. Are therefore GRGs just evolved FR\,II galaxies, for which large linear sizes are strictly due to their advanced ages of the order of $100$\,Myr? Or are they instead intrinsically different from classical doubles \citep[as suggested by, e.g.,][]{kai99}? A detailed analysis of various samples of GRGs favors the former possibility, suggesting however in addition that densities of the medium surrounding the discussed objects are lower than average for the whole FR\,II population \citep{mack98,sch00b,mach06}. It is interesting to note in this context an analogy with the sources located at the other extremum of the size distribution of radio galaxies, namely Compact Symmetric Objects (CSOs), which spectroscopically are often classified as GigaHertz-Peaked Spectrum (GPS) sources. For these, the long-debated question was if their linear sizes $0.1 - 10$\,kpc are due to young ages ($\leq 0.1$\,Myr) of otherwise `regular' FR\,II-like radio structures, or instead are due to exceptionally dense ambient medium frustrating evolved jets and confining their lobes within host galaxies \citep[see][for a review]{odea98}. Or do they constitute yet another intrinsically distinct population of extragalactic radio sources? The merging agreement is that compact radio galaxies are indeed newly born precursors of classical doubles \citep[and also of low-power FR\,I sources; see the discussion in][and references therein]{sta08}, even though some of them may reside in particularly overdense environment \citep[see, e.g.,][]{gar07}.

During the recent years it became clear that understanding the evolution of extragalactic radio sources --- including the youngest and the oldest examples of this class of astrophysical objects --- involves however not only accessing the properties of the ambient medium into which they evolve, or the main jet parameters (like jet total kinetic power, lifetime, etc.), but also characterizing a complex duty cycle of the jet activity. In particular, population studies of radio galaxies and quasars indicated that the jet activity must be intermittent, or at least highly modulated, in addition on different timescales \citep[e.g.,][]{rey97,alex00,sch01,cze09}. The direct observational evidences for such an intermittency consist of different manifestations of the interrupted and/or restarted jets in radio-loud Active Galactic Nuclei (AGN), including interestingly some GRGs with compact inner structures of the CSO/GPS type \citep[e.g.,][]{sch99,kon04,kon08,sai07}. This is an important point indeed, since it indicates that the appearance of large-scale jets and lobes is shaped not exclusively by the jet lifetime and the jet kinetic power relative to the density of the ambient medium, as believed previously, but is a complex convolution of these factors with the jet duty cycle. Note in particular that the restarted jets do not propagate within the undisturbed interstellar/intergalactic medium, but instead within the environment modified by the passage of the outflow during the previous epoch of the jet activity, and this may affect substantially the observed properties of the newly formed lobes.

Double-double radio galaxies (hereafter referred to as DDRGs) are the most striking and clear examples of a highly intermittent activity of the jet engine with no realignment of the outflows' direction in the two subsequent activity epochs \citep{lar99,sch00a}. These are the sources in which an extended radio structure comprises at least two pairs of lobes with a common core, both strongly co-aligned to within a few degrees to each other. At present, about 15 of such systems have been identified from either radio or X-ray observations \citep[for a review see][]{sai09}. The outer and also the inner lobes of DDRGs are uniquely characterized by an edge-brightened morphology of the FR\,II type, what led at the first place to the interpretation of the inner structures in DDRGs as `regular lobes' indeed, inflated by a plasma left behind a head (a termination shock) of a new-born jet which propagates through an outer cocoon formed during the previous epoch of the jet activity \citep{kai00}. The alternative interpretation for the inner structures of DDRGs would involve ballistic and `naked' (lobeless) jets observed in a few performed 2D and 3D hydrodynamical simulations of restarting outflows in AGN \citep{cla91,cla97}. Bulk of the observations and analysis carried till now for several best known DDRGs are in general consistent with the former scenario \citep[see, e.g.,][]{sar02,sar03,kon06,saf08,mach10}, even though some potential problems of the model has been recently discussed in \citet{bro07,bro11}.

The problems arise when the exact evolution of the inner lobes of DDRGs is analyzed. These inner lobes appear rather slim on the high-dynamics radio maps, displaying in particular axial ratios in the range of $9 - 16$, to be compared with the range of $2 - 5$ established for the outer lobes of same systems, and also for the lobes in `regular' FR\,II radio galaxies. This indicates that the restarted jets in DDRGs propagate within the environment much rarified with respect to the original IGM in which the outer lobes are evolving, as in fact expected in the framework of the jet intermittency scenario, but yet much denser than the one provided simply by the injection of a jet plasma to the outer lobes in the previous epoch of the jet activity \citep[see, e.g., the detailed analysis by][]{saf08}. Thus required contamination of the remnant cocoons by the additional matter is one of the central questions regarding the nature of DDRGs \citep[see the discussion in][]{kai00}. It is not clear, however, if any entrainment process operates on sufficiently short timescales to enrich substantially the outer cocoons with the required amount of the cold matter from the IGM during the lifetime of a radio source. For this reason, \citet{bro07,bro11} argued that radio emission observed from the inner structures of DDRGs is not due to the lobes of newly born jets, but predominantly due to bow shocks induced within the extended lobes by only briefly interrupted relativistic outflows. In this scenario no contamination of the extended/outer cocoon by the additional matter would be needed.

In this paper we consider and analyze a possibility that the structure of the largest radio galaxy known up to date --- \J with the total projected linear size of $D=4.69$\,Mpc, discovered by \citet{mach08} --- is formed by a restarted, rather than a primary jet activity. The motivation for the performed analysis and new radio observations is due to the fact that the axial ratio for the lobes in this outstanding example of a giant, $R_{\rm T}\approx 12$, is of the same order or even larger than axial ratios for the inner lobes in all well studied DDRGs, and therefore very different from the typical axial ratios characterizing `normal size' FR\,II sources or other known GRGs. Since in the framework of the standard dynamical model for classical doubles the axial ratio of the lobes depends predominantly on the jet kinetic power relative to the density of the surrounding, \J must be a powerful source evolving in a particularly low-density environment, as indeed concluded in \citet{mach08}. The question is if the low-density of the matter surrounding giant lobes of \J is due to a unique location of \J in a void region, or due to some previous jet activity epoch which took place before the formation of the currently observed lobes, and which rarified the IGM thereby (in analogy with DDRGs). In the former case, \J may be used as a particularly interesting (due to its enormous size) probe of a `warm-hot phase of the intergalactic medium' (hereafter WHIM) within a filamentary galaxy distribution far away from any group or cluster of galaxies. In the latter case, \J would be the most extreme example of a restarting radio galaxy, providing interesting constraints on the jet duty cycle and cosmological evolution of radio-loud AGN.

New observations of \J, conducted with the Giant Metrewave Radio Telescope (GMRT) and Very Large Array (VLA), confirm its morphology consisting of only two narrow lobes discerned in the VLA $1400$\,MHz sky surveys FIRST \citep{bec95} and NVSS \citep{con98}, together with the radio core coincident with a parent galaxy at the redshift of $z=0.3067$. The whole radio structure is highly collinear and symmetric: the arm ratio is 1.08 only, and the misalignment angle is $1.3 \degr$. No trace of any outer cocoon is however detected in our new observations. We modeled \J using the \D algorithm of \citet{mach07}, along with the three bona-fide DDRGs treated as comparison and control sources. The resulting `radio luminosity---linear size' $(P_{\nu}-D)$ evolutionary diagrams are used to determine conditions allowing the hypothetical outer lobes in the discussed object to dim in radio fluxes to a level precluding the detection at the given sensitivity limits. The archival data regarding the exemplary DDRGs used in the modeling, as well as the new GMRT and VLA observations of \J resulting in high-resolution low- and high-frequency maps of the source, are presented in \S\,2 of the paper. The applied model and the obtained model results are described in \S\,3. Further discussion regarding the jet energetics and the properties of the environment of \J are given in \S\,4. Final conclusions are summarized in \S\,5. Distances, linear sizes, volumes, and luminosities of the analyzed sources are calculated for the modern $\Lambda$CDM cosmology ($\Omega_{\rm m}=0.27$, $\Omega_{\Lambda}=0.73$, and $H_{0}=71$\,km\,s$^{-1}$\,Mpc$^{-1}$).

\section{The Data}

In order to test the hypothesis that the observed radio structure of \J results from a secondary jet activity in the parent AGN, we gathered new GMRT and VLA data. These allow us to model the source using the \D algorithm (see \S\,3.1), along with the three carefully selected DDRGs, treated as comparison sources, for which the outer and inner lobes are observed and well mapped at a number of radio frequencies. The results of the newly conducted observations of \J as well as the details regarding the selected DDRGs are given below. Main properties of the studied sources are summarized in Table\,1.

\subsection{GMRT and VLA Observations of \J}

The analysis of \J presented in this paper is based on the radio observations conducted with the GMRT and VLA, as well as on the archival VLA data from the NRAO $1400$\,MHz FIRST and NVSS surveys \citep{bec95,con98}. The observing log for the new GMRT and VLA observations is given in Table\,2, including the name of the telescope and the array configuration, the frequency of observations and the primary beamwidth, as well as the dates of the exposures. The $4860.1$\,MHz VLA image of the entire source is presented in Figure\,1. The enlarged GMRT and VLA images of the northern and southern lobes at four different observing frequencies are shown in Figure\,2. The images at $1400$\,MHz are reproduced from the FIRST survey, and the instrumental characteristics for all the radio maps included in this paper are collected in Table\,3. The obtained radio fluxes of \J and the resulting luminosities are listed in Table\,4.

The low-frequency GMRT observations at $328.8$\,MHz and $617.3$\,MHz were made in the standard manner, with each observation of the target source interspersed with observations of calibrator sources. The phase calibrators PKS B\,1416$+$067 (3C\,298) and PKS B\,1442$+$101 were observed at $328.8$\,MHz and $617.3$\,MHz, respectively, after each of several 20 minute-long exposures of the target centered on the core position. 3C\,286 was used as the flux density and bandpass calibrator based on the scale of \citet{baa77}. The total observing times on the target sources were about 200 min (at $328.8$\,MHz) and 180 min (at $617.3$\,MHz). The data were edited and reduced using the NRAO \A package, and then self-calibrated to produce the best possible images. 

The high-frequency $4860.1$\,MHz observations conducted with the VLA in its D configuration were made in two runs aimed at a different lobe of \J each (see Table\,2). The interferometric phases were calibrated every 20 min with the phase calibrator PKS B\,1351$-$018. The source 3C\,286 was used as the primary flux density calibrator. 41 and 93 minute-long exposures were conducted for the fields centered on the northern and the southern lobes, respectively. This allowed us to reach an rms noise values of about 0.048 and 0.036\,mJy\,beam$^{-1}$ for the two exposures respectively. The data were edited, reduced and self-calibrated with the NRAO \A package. Images produced from the resulting datasets were corrected for the primary beam pattern. 

\subsection{Comparison Sample of DDRGs}

The three DDRGs selected for this project are J1453+3308, J1548$-$3216, and J0041+3224. The archival data for these used in our modeling are summarized in Table\,5.

The source J1453+3308 ($\equiv$ B1450+333) is one of the best studied DDRG \citep{sch00a,kai00,kon06}. Multifrequency radio maps of this object presented in \citet{kon06} show broad, diffuse outer lobes extended out of the collinear pair of narrow inner lobes. The dynamical modeling of the source discussed in \citet{mach09} and in \citet{bro11} resulted in significantly different values for the fitted model parameters. This is most likely due to different model setups and different assumptions involved in the modeling. Hence J1453+3308 deserves further consideration.

The source J1543$-$3216 ($\equiv$ PKS B1545$-$321) was analyzed by \citet{sar03}, \citet{saf08}, and \citet{mach10}. The outer lobes of this DDRG extend back to the parent galaxy forming a common cocoon with no significant gaps in radio brightness. The inner narrow lobes are therefore fully embedded within the outer lobes, all the way from the radio core to their termination points. The ages and other characteristics of both pairs of lobes, as well as the properties of their surrounding environments, have been previously determined using the \D algorithm by \citet{mach10}.

The DDRG J0041+3224 was discovered by \citet{sai06}. In spite of the lack of spectroscopic redshift determination for the parent galaxy, the source is taken into consideration in our work because its inner lobes are brighter at radio wavelengths than the outer lobes. This is quite unusual characteristic in the whole sample of all known DDRGs. The other morphological properties of the source are however typical for its class. An uncertainty in the photometric redshift estimate for J0041+3224 affects the determination of its physical size and luminosity, but should not influence much the main conclusions.

\section{Modeling}

The standard and widely considered scenario for the time evolution of FR\,II type radio sources is based on the pioneering work by \citet{sch74} and \citet{bla74}, modified subsequently by a number of authors \citep[e.g.,][]{beg89,fal91,kai97,KDA97,blu99,man02}. This scenario enables for a self-consistent evaluation of the source (lobes') length and radio luminosity for a given jet power, jet lifetime, and density of the ambient medium, accounting for the cumulative effects of energy losses of relativistic particles injected by the jet into the evolving cocoon due to the synchrotron radiation, inverse Compton scattering of CMB photons, and adiabatic expansion of the lobes. Our analysis presented below is based on a particular dynamical description by \citet{kai97} with the radiative processes properly taken into account following Kaiser, Dennett-Thorpe \& Alexander (1997; hereafter `KDA' model), as summarized in \citet{kai07}.

\subsection{\D Algorithm}

Similarly to our previous work \citep{mach08,mach09,mach10}, the analysis presented in this paper is performed using the \D algorithm of \citet{mach07}. This numerical code allows us to solve the inverse problem, i.e. to determine --- for a given set of observables --- four main free parameters of the KDA model, namely (i) the jet power $Q_{\rm j}$, (ii) the density of the ambient gaseous medium near the radio core $\rho_{0}$, (iii) the age of the lobes $t$, and (iv) the \emph{effective} injection spectral index $\alpha_{\rm inj}$. The latter parameter approximates the slope of the emission continuum of the electron energy distribution freshly injected at the jet head --- jet termination shock (hotspot) --- into the expanding lobes. Note that the term `effective' stands here for the average over a broad energy range and over the lifetime of the source, since the energy spectrum of particles accelerated at the hotspot may change with time, and may be far from a `single power-law' approximation \citep[see the discussion in][]{mach07}. A relatively robust determination of these four free parameters is enabled by fitting the applied dynamical/radiative model to the main observables, namely (i) the projected linear length of the cocoon $D$, (ii) the total volume of the cocoon $V$, and (iii) the lobes' radio power at different frequencies $P_{\nu_{i}}$ (where $i=1, 2, 3,. . .$). As in the KDA model, we assume an inclination $\theta = 90\degr$ and a cylindrical geometry for each source. This, for the total length $D$ and width $w$, gives the total volume $V=\pi D^{3}/(4\,AR^{2})$, where $AR\equiv 2\,R_{\rm T}=D/w$ is the axial ratio of the whole radio structure (in the case of roughly symmetrical lobes). The other model parameters (such as adiabatic indices for the plasma inside and outside of the cocoon, $\hat{\gamma}_{\rm c}$ and $\hat{\gamma}_{\rm IGM}$, respectively) and model assumptions are discussed in \citet{mach07}.

The original KDA model assumes a constant and uninterrupted jet flow delivering ultrarelativistic gas to the expanding lobes throughout the whole lifetime of a source. As such, it is well suited for describing relatively young or normal-size FR\,II radio galaxies, but may not be directly applicable for modeling the evolved objects, i.e. remnant/outer lobes of GRGs and DDRGs, whose observed properties indicate in many cases a faded jet activity at the time of observations. Therefore, in order to analyze such systems in the framework of the discussed scenario, it seems necessary to introduce one more free parameter to the model, namely the jet lifetime $t_{\rm j}$, which may be substantially different than the age of the source $t$. Note in this context that when the jet activity is terminated, the adiabatic expansion of the cocoon inflated by the jets is expected to slow down, or even to stop completely \citep[see][]{bro07,bro11}. Hence two different phases of the dynamical evolution of such lobes should in fact be considered, the first one corresponding to the ongoing feeding of the lobes by the active jets (source ages $t \leq t_{\rm j}$), and the second one corresponding to the case when the jet flow is ceased ($t>t_{\rm j}$). As discussed in \citet{bro11}, a few/several sound-crossing times through the lobes are needed to establish the new expansion regime during the second phase of the evolution, and this timescale may be comparable to the age of the source in the case of large-scale outer lobes of GRGs and DDRGs. 

Following the above argument, in our fitting procedure we neglect the effect of a slower expansion phase when evaluating the dynamical parameters of a source. This allows us in particular to estimate the length of a jet as
\begin{equation}
\ell_{\rm j}(t)=c_{1}\left(\frac{Q_{\rm j}}{\rho_{0}\,a_{0}^{\beta}}\right)^{1/(5-\beta)}t^{3/(5-\beta)} \, ,
\end{equation}
\noindent
and the longitudinal expansion velocity of a cocoon as
\begin{equation}
v_{\rm h}(t)=\frac{{\rm d}\ell_{\rm j}}{{\rm d}t}=\frac{3\,c_{1}}{5-\beta}\left(\frac{Q_{\rm j}}{\rho_{0}\,a_{0}^{\beta}}
\right)^{1/(5-\beta)}t^{(\beta-2)/(5-\beta)} \, .
\end{equation}
\noindent
In the above $c_{1}$ is a numerical constant depending on a value of $R_{\rm T}$, $a_{0}$ is the radius of the central plateau in the density distribution of the surrounding gas, and $\beta$ is the exponent in the anticipated ambient density profile $\rho(r)=\rho_{0}\,(r/a_{0})^{-\beta}$. The total projected length of the entire structure is $D=2\, \ell_{\rm j}$. The ratio of the pressure within the lobes' heads to that within the rest of the cocoon, ${\cal P}_{\rm hc}\equiv p_{\rm h}/p_{\rm c}$, is taken as ${\cal P}_{\rm hc}=(2.14-0.52\,\beta)\,R_{\rm T}^{2.04-0.25\,\beta}$ following \citet{k00}. 

We note that only if $t_{\rm j}$ is \emph{significantly} shorter than $t$, thus evaluated model parameters would correspond to the \emph{lower limit} on the ratio $Q_{\rm j}/\rho_{0}$. But since in our discussion below (\S\,4) we emphasize large jet kinetic powers and low ambient medium densities required to fit the observed spectrum of \J (and of the other giant sources), equations 1--2 constitute a very conservative choice indeed. Besides, in the presented model fitting the jet lifetimes are never less than $70\%$ of the source ages. Importantly, the evaluation of the lobes' luminosities is done taking into account only those relativistic particles which are injected into the lobes during the first phase of the evolution, corresponding to the ongoing jet activity. The integration over the source spectrum is performed following the description provided in \citet{kai02} in the context of relict radio sources. 

In the KDA model the minimum-energy condition is applied, being determined by accounting for the energies stored in the lobes' magnetic field and in relativistic radiating particles only. However, the extended lobes of luminous radio galaxies may not fulfill exactly the minimum-energy condition, and may contain --- especially at large stages of their evolution --- a significant fraction of non-radiating (possibly mildly-relativistic) particles, as indicated by the recent X-ray observations of a number of such systems as well as by some theoretical considerations. Yet the departures from the energy equipartition between magnetic field and radiating electrons within the FR\,II lobes are never that large \citep{kat05,cro05}, as demonstrated also for several particular cases of GRGs \citep{kon09,iso09,iso11a,iso11b}. Hence, in our analysis we apply the standard minimum-energy condition, but allow for a non-negligible contribution of non-radiating particles (ultrarelativistic or only mildly-relativistic protons) to the total pressure inside the cocoons. The ratio of the energy density stored in the lobes' magnetic field $B$ to that stored in all the lobes' particles is then ${\cal R}_{\rm eq}=(1+s_{\rm inj})/[4\,(k^{\prime}+1)]$, where $s_{\rm inj}=2\,\alpha_{\rm inj}+1$ is the energy slope of the electrons (with minimum and maximum Lorentz factors $\gamma_{\rm min}$ and $\gamma_{\rm min}$, respectively) freshly accelerated at the jet termination shock, and $k^{\prime}$ is the energy/pressure fraction of the non-radiating particles.

\subsection{Results}

The observational data summarized in \S\,2 are fitted using the \D algorithm described in \S\,3.1. The results of the modeling are given in Table\,6. In the fitting procedure we fix the following model parameters: adiabatic indices for the plasma within and outside of the cocoons $\hat{\gamma}_{\rm c} = \hat{\gamma}_{\rm IGM} = 5/3$, minimum electron Lorentz factor $\gamma_{\rm min} = 1$ and the maximum electron Lorentz factor $\gamma_{\rm max} = 10^7$, as well as the inclination of the lobes $\theta = 90\degr$ \citep[see][for more discussion regarding these model assumption]{mach07}. In the anticipated density profile characterizing the ambient medium into which the lobes evolve, $\rho(r)=\rho_{0}\,(r/a_{0})^{-\beta}$, we fix the characteristic radius of the density plateau $a_0 = 10$\,kpc, and --- in the case of the outer cocoons propagating through an undisturbed IGM --- the power-law index $\beta = 1.5$. Such a density profile is expected to be a good approximation of the hot gaseous medium in groups of galaxies hosting luminous radio sources \citep{KDA97}. When modeling the inner lobes, which propagate within a uniform plasma of the outer cocoons, we set instead $\beta = 0$ or 0.1 (as determined by a quality of the fits). Note that the assumption regarding a uniform distribution of the matter within the lobes is justified by the fact that the sound speed thereby has to be in general high when compared to the dynamical timescale involved \citep[see][]{kai00}. Finally, we assumed the fraction of the non-radiating particles $k^{\prime} = 0$ for the inner lobes, and $k^{\prime} = 10$ for the outer lobes, in order to account for the expected entrainment processes enriching the evolved cocoons with the matter.

The model fits return the following free parameters: the total jet kinetic power $Q_{\rm j}$, the density of the central plateau in the distribution of the ambient medium $\rho_{0}$, the injection spectral index $\alpha_{\rm inj}$, the age of the lobes $t$, and the jet lifetime $t_{\rm j}$. These allow us to evaluate further some other parameters of the analyzed sources, including the advance velocity of the jets $v_{\rm h}(t)/c$ (see equation\,2), the average expansion velocity of the source $D/(2 c\,t)$, the density of the surrounding medium around the lobes' heads $\rho(D)$, the cocoon pressure $p_{\rm c}$, and the magnetic field intensity within the lobes $B$. The radio spectra for the control sample of DDRGs fitted by the applied model are presented in Figure\,3. The quality of the fits is quantified by the reduced $\chi^{2}$ values given in Table\,6, and also in Figure\,3. 

\subsection{The Case of J1420$-$0545}

The high-resolution GMRT image of \J at 617\,MHz confirms very narrow lobes of the source, as already suggested by the 1400\,MHz map available from the FIRST survey. The axial ratio of the lobes, $R_{\rm T}\approx 12$, is of the same order or even bit higher than that characterizing the inner lobes in basically all the well-studied DDRGs, and much higher than that established for normal-size FR\,IIs, outer lobes of DDRGs, or GRGs. Also the age of the observed radio structure of \J found from the preliminary model fitting presented by \citet{mach08}, namely $\simeq 50$\,Myr, as well as the average expansion speed evaluated by the authors, $\gtrsim 0.1\,c$, are totally discordant with the corresponding values determined for the other GRGs. Therefore, the idea that the observed structure of \J might be due to a restarted rather than a primary jet activity seems well justified. Unfortunately, all the observations made till now do not show a trace of any relict (older) cocoon embedding the observed lobes.

In the first step of the modeling, we repeated the calculations of \citet{mach08} using the new radio data presented in \S\,2. The observed lobes of \J are thus modeled as structures formed by a primary jet activity, with the standard ambient medium density profile appropriate for group of galaxies assumed ($a_0 = 10$\,kpc and $\beta = 1.5$), along with $k^{\prime} = 1$ (model ``Primary'' in Table\,6). It is worth noting however that the assumption regarding the ambient medium density may not be strictly valid in the particular case of the analyzed object, since the optical galaxy counts around \J do not reveal any excess over the background level (see in this context \S\,4.1). Besides, even if some gaseous halo exists around the host galaxy of the source, similar to the virialized hot IGM of elliptical galaxies and groups of galaxies, its extension up to $>$\,Mpc distances from the central AGN in a form of a steep power-law density distribution seems questionable. Thus, the parameters returned by the ``Primary'' model fit have to be taken with caution. In addition to this, we also model the lobes of \J as structures inflated by restarting jets, in analogy to the inner lobes of DDRGs, setting in this case $\beta = 0.1$ and $k^{\prime} = 0$ (model ``Inner'' in Table\,6). For illustration, the ``Inner'' model fit to the comoving spectrum of the entire structure of \J is shown in Figure\,4.

Both ``Primary'' and ``Inner'' models return comparable physical parameters of the studied object --- in particular $Q_{\rm j} \simeq 4 \times 10^{45}$\,erg\,s$^{-1}$, $\alpha_{\rm inj} \simeq 0.55$, $t \simeq 35$\,Myr, $v_{\rm h}(t)/c \simeq D/(2 c\,t) \simeq 0.2$, $p_{\rm c} \simeq (0.1-0.4) \times 10^{-13}$\,dyn\,cm$^{-2}$, and $B \simeq 0.6$\,$\mu$G --- except of the central density of the ambient medium $\rho_{0}$. We note here relatively young age of the source implied by the model fits (in between of the ages evaluated for the inner and outer lobes of the analyzed DDRGs), its relatively high expansion velocity (ten times higher that the corresponding values found for the outer lobes in the control sample of DDRGs), and the magnetic field intensity exactly within the range established for other GRGs through the X-ray observations \citep{kon09,iso09,iso11a,iso11b}. The goodness of the fits in both ``Primary'' and ``Inner'' cases is practically the same; the test $\chi^{2}_{\rm red}$ cannot be therefore used to distinguish between the models. Importantly, the only difference between the two models considered, i.e. the emerging value of the $\rho_0$ parameter, is just a direct consequence of the different ambient density profiles \emph{assumed} ($\beta = 1.5$ versus $\beta =0.1$); indeed, the density of the surrounding medium \emph{around} the lobes' heads, which is of a main importance in determining the other model parameter for the evolved/giant system like \J, is basically identical in both scenarios, and is in addition \emph{very low}, namely $\rho(D) \simeq 2 \times 10^{-31}$\,g\,cm$^{-3}$. Hence the emerging conclusion is that regardless on the unknown density profile of the ambient medium surrounding the giant lobes of \J and regardless on the unknown contribution from the non-radiating particles to the internal pressure of its cocoon (at least within the explored range) the lobes of \J must evolve in a very rarified surrounding. We return to this point in the next section \S\,4.

If the low density medium surrounding \J is due to the previous jet activity, i.e., if this source is indeed of a double-double type, the question is where are its outer lobes. As already emphasized above, such structures are not detected in the newly performed radio observations, but this may be only because of their advanced ages precluding any detection due to substantial cooling of the radiating electrons. In order to explore this idea in more detail, we evaluate the expected properties of the hypothetical outer lobes, within the framework of the general model for the evolution of luminous radio galaxies anticipated in this paper, by assuming the jet power in the previous activity epoch three times larger than the power emerging from the ``Inner'' modeling of \J, namely $Q_{\rm j} \simeq 1.2 \times 10^{46}$\,erg\,s$^{-1}$. This is in analogy to the control sample of DDRGs, for which the previous jet activity epochs seem to be universally characterized by larger kinetic powers of the outflows when compared to the ongoing jet activity epochs. We also fix the ambient medium density profile $\beta = 1.5$, and set the central density as $\rho_{0} = 4 \times 10^{-26}$\,g\,cm$^{-3}$, which is the average determined for 30 GRGs studied by \citet{mach09} and \citet{mach11}. Finally, we set the aspect ratio for the hypothetical relict lobes of \J as half of the value characterizing the observed radio structure, $R_{\rm T}=6$. The expected time evolution of the length and luminosity of the outer lobes is then calculated in a number of models differentiated by values of $t_{\rm j}$ and $t$. Searching for a model providing relict lobes' length not shorter than that of the currently observed structure, as well as radio fluxes below the current sensitivity limits, we find a possible solution for $D = 4.70$\,Mpc, $t_{\rm j}=106$\,Myr and $t_{\rm out}=178$\,Myr (see model ``Outer'' in Table\,6). Figure\,5 presents the expected observed spectrum of the hypothetical outer lobes corresponding to the model solution.

\subsection{Evolutionary Tracks}

In the framework of the considered model and for a given set of the observables, the whole time evolution can be easily calculated for each analyzed source. Figure\,6 shows such an evolution, depicted in the luminosity--size evolutionary tracks, $P_{\nu} - D$, for the outer and inner lobes of the three exemplary DDRGs modeled in this paper. Here the star symbols indicate the actual ages of the lobes emerging from the modeling, while the consecutive age markers are connected with dotted lines. The diagrams indicate that very soon after a termination of the jet flow the outer lobes decrease quickly in radio luminosity, and slow down their expansion. The inner lobes may then `overtake' the outer cocoon both in the luminosity and in size. Strictly speaking, at the point when the new-born structures grow beyond the older ones the model applied here to describe the evolution of the inner lobes is not valid anymore; hence the segments of the evolutionary curves shown in Figure\,6 which corresponds to such a case do not describe any realistic situation. However, our calculations indicate also that this is expected to happen only when the inner lobes reach an age of about or over 100\,Myr. 

The analogous evolutionary tracks for \J are shown in Figure\,7. The tracks corresponding to the ``Primary'' and ``Inner'' models for the observed radio structure are denoted by solid and dashed curves, respectively. The evolutionary track for the hypothetical outer lobes in \J, corresponding to the ``Outer'' model parameters, is given by short-dashed curve. The actual ages of the observed and of the hypothetical/outer radio structures emerging from the model fits are marked by the filled and open star symbols, respectively.

As can be inferred from the presented diagrams 6--7, for the ages of the outer lobes $t<t_{\rm j}$ corresponding to $D(t) \geq a_{0}$ the lobes' luminosities decrease slowly with time. This is because of the ongoing synchrotron and adiabatic cooling of ultrarelativistic electrons evolving within slowly expanding cocoons. We note that in the framework of the applied model the adiabatic losses do not depend on the age of a source, due to the assumed constant ratio between the lobes' pressures ${\cal P}_{\rm hc}\equiv p_{\rm h}/p_{\rm c}$ throughout the lifetime of a system. Also, inverse-Compton energy losses of the radiating particles are mostly negligible: for all the outer lobes modeled in this paper the energy densities stored in the lobes' magnetic field exceed the energy densities of the CMB photon field at the source redshifts, $u_{\rm B} \equiv B^2/8\pi > u_{\rm CMB} \simeq 4 \times 10^{-13} (1+z)^4$\,erg\,cm$^{-3}$, up to late phases of the evolution $t \gtrsim 100$\,Myr. On the other hand, once the jet flow decays and the source grow old (ages $t>t_{\rm j}$), combined synchrotron and inverse-Compton energy losses quickly cool most of the particles radiating within the relict lobes, causing a rapid decrease in the source luminosity.

The inner lobes evolve in a different manner. In the uniform, low-density but high-pressure environment their adiabatic (sideway) expansion is hampered, and as a result the lobes' luminosity increases with time. However, the simultaneously decreasing energy density of the lobes' magnetic field falls at some point below the energy density of the CMB photon field, causing the onset of the luminosity decline due to the dominant inverse-Compton energy losses of the radiating electrons. It is worth noting that such an effect, predicted already by \citet{bal82}, strongly depends on the density of the source environment. This is illustrated in Figure\,8, which shows the $P_{\nu} - D$ evolutionary tracks of a fiducial source \citep[as considered by][]{KDA97} evolving within a uniform ambient medium ($\beta$=0) characterized by three different values of the density: $\rho_{0} = 7.2 \times 10^{-25}$\,g\,cm$^{-3}$ \citep[solid line, reproducing figure\,4 in][]{KDA97}, $5 \times 10^{-26}$\,g\,cm$^{-3}$ (dashed line), and $5 \times 10^{-28}$\,g\,cm$^{-3}$ (dotted line). As illustrated, the effect is stronger and stronger for lower and lower density of the ambient medium, but in all the cases the inverse-Compton dominance and the related onset of the luminosity decline starts around the source ages of about $t \lesssim 100$\,Myr, when the source sizes are $D \sim 0.1-1$\,Mpc. The time evolution of length and
luminosity of the outer and inner lobes of the exemplary DDRG J1548$-$3216 is shown in Figure\,9.

\section{Discussion}

The main observational argument for a restarted jet activity in \J is the very high axial ratio of the observed radio structure. This finding suggests almost ballistic propagation of powerful jets in a particularly low-density ambient medium. If this low density of the source environment is due to the previous jet activity indeed, or rather due to a unique location of \J in a void region of the galaxy distribution, is a central question we aim to address in the presented analysis. Extremely large linear size of the studied object, $D \simeq 4.7$\,Mpc, makes the investigation particularly relevant. Note that radio cocoons exceeding a few Mpc in size are quite unexpected at the redshift of 0.3, since all the other extended FR\,II systems at $z \leq 0.3$, including outer lobes of DDRGs, are substantially shorter than this. Regardless on the details of the jet duty cycle and of the origin of the lobes in the system, however, the energetic requirements for the outflows in \J are severe, as discussed below.

\subsection{Galaxy Counts} 

As an independent check of the IGM density in the vicinity of \J we have performed simple galaxy counts around the position of the source, and compared the resulting values with the mean `background' number density of galaxies determined by \citet{tys88},
\begin{equation}
\log\,{\cal N}_{\rm bg} = \left( 0.39 \times R - 4.8 \right) \,{\rm deg^{-2}\,mag^{-1}} \, ,
\end{equation}
\noindent
for the isophotal red magnitudes $R$. We note that the isophotal limits are sufficiently low for most of the galaxies so that $R$ can be considered as total magnitudes to the precision of the faint-galaxy photometry. However, in order to perform reliable counts of galaxies surrounding the analyzed target, it is necessary to choose proper selection criteria in order to remove from the list stellar objects, keeping in mind that the host galaxy of \J itself is very faint: $R=19.65$\,mag and $B=21.82$\,mag \citep{mach08}.

Luminous extragalactic radio sources are hosted by elliptical galaxies located in the centers of poor clusters and groups. A standard measure of cluster richness can therefore be used to check whether \J shares this characteristic or nor. Here we follow a particularly simple and frequently applied procedure of counting the objects within 3\,Mpc radius around a given $R^{\star}$--magnitude galaxy, which are brighter than $R^{\star}+3$ \cite[e.g.,][]{hil91}. For this purpose we use the Minnesota Automated Plate Scanner (MAPS) Catalog of the first-epoch Palomar Observatory Sky Survey (POSS-I), which was derived from digitized scans of glass copies of the blue and red plates of the original POSS-I. The limiting magnitudes of this database preclude however detections of the systems as faint as $R = 19.65+3 = 22.65$\,mag. For this reason, we define ${\cal N}$ as a number of galaxies with red POSS-I magnitudes $E \pm 1.5$, which are located within 3\,Mpc radius around \J (corresponding to the angular separation of $11.1\,\arcmin$). Two different selection criteria have been considered: (i) $0.5\leq{\rm galnod}\leq 1.0$ at both $E$ and $O$ plates indexing non-stellar objects, and (ii) colors $O-E>1.4$\,mag providing rejection of bluer objects, i.e., by assumption, quasars. As a result, 68 and 72 objects fulfilling the above criteria have been found, respectively. The emerging numbers are, quite surprisingly, fully compatible with each other. Diving the corresponding mean value $70$ by a circular sky area of 3\,Mpc radius at the position of the target source gives us ${\cal N} \simeq 650$ galaxies per deg$^{2}$. The relevant number of `background' galaxies implied by equation\,3 is ${\cal N}_{\rm bg} \simeq 2650$. Hence, the resulting ratio ${\cal N}/{\cal N}_{\rm bg} \simeq 0.25$ indicates that \J is indeed located in distinctly poor region of the galaxy distribution, as further confirmed by investigating the number density of galaxies within a few larger concentric rings, up to 16\,Mpc from the target; this number density occurs to be almost constant.

\subsection{Relict Lobes?} 

The $P_{\nu}-D$ diagrams shown in Figure\,6 illustrate that after a termination of the jet activity, luminosities of the outer lobes in DDRGs decrease quickly with time, falling eventually below the inner lobes' luminosities. This is further depicted in Figure\,9 for the particular case of DDRG J1548$-$3216. Therefore, if the hypothetical outer lobes of \J are old enough, one may expect their radio surface brightness to fall substantially below the rms noise level of the available radio maps. Figure\,7 shows that the 1400\,MHz luminosity of these lobes, calculated for the ``Outer'' model parameters (Table\,6), is indeed lower than that of the currently observed structure in the system. On the other hand, one should expect at the same time a substantial steepening of the radio continuum for the dying cocoon, and hence possibly non-negligible fluxes at low observing frequencies. Figure\,5 indicates in particular that the total flux density of the presumed outer lobes in \J might be comparable to the observed flux density at frequencies $\lesssim 300$\,MHz. Around 74\,MHz, the frequency of the VLSS survey \citep{coh07}, the estimated flux is about 1.4\,Jy. Even though relatively high, this flux is still not enough for a robust detection of the relicts in the archival dataset. We note that the VLSS survey is characterized by the average sensitivity of $\sigma \sim 0.1$\,Jy per beam of $80\arcsec\times 80\arcsec$, and hence that the detection limit of about $5\sigma$ for an unresolved source reads as $\sim 0.5$\,Jy\,beam$^{-1}$. Assuming therefore that each of the two putative outer lobes of \J has an area of about $2\arcmin\times 3\arcmin$, i.e. about $3 -4$ beams, a 74\,MHz flux of about $1.5 -2.0$\,Jy would be required for the detection in the VLSS. This is $2-3$ times the $\sim 0.7$\,Jy flux expected for each of the lobes in the framework of the ``Outer'' model. 

Future high-sensitivity VLA or Low Frequency Array (LOFAR) observations at frequencies below 100\,MHz could be however more promising for investigating a presence of the hypothetical outer cocoon in \J. Interestingly, this structure could be also possibly probed at X-ray frequencies, since inverse-Compton scattering of the CMB photons to the X-ray range involves electron energies lower than those involved in producing observable radio synchrotron emission of the lobes. Searching for analogous relict (`ghost') structures in other evolved radio galaxies and radio-loud quasars using current X-ray instruments has been in fact already proposed and considered in the literature \citep[e.g.,][]{fab09,blu11,moc11}.

\subsection{Age and Expansion Velocity}

In the previous section \S\,3.3 we have noted a relatively young age of the observed radio structure of \J emerging from the model fits, $t \simeq 35$\,Myr, which is in between of the ages evaluated for the inner and outer lobes of the analyzed DDRGs. Correspondingly, a relatively high expansion velocity $v_{\rm h}(t)/c \simeq D/(2 c\,t) \simeq 0.2$ has been found for the source, about ten times higher that the analogous values estimated for the outer lobes in the control sample of DDRGs, and a factor of at least a few higher than the expansion velocity of the inner lobes in such systems. Are however these somehow extreme values realistic at all, or are they just artifacts of the incorrect scenario applied? 

In all the kinematic models for FR\,II-type sources, characteristic asymmetries in lengths and luminosities between the two opposite lobes are expected due to the light time travel effects \citep[e.g.,][]{lon79,bes95,sch95,rys00}. The size ratio between the longer (jet side) and shorter (counter-jet side) lobes is in particular $q=[1+(v_{\rm h}/c) \cos\theta]/[1-(v_{\rm h}/c) \cos\theta]$, giving the advance speed of the jets
\begin{equation}
v_{\rm h} = \left( \frac{q-1}{q+1} \right) \, \left(\frac{c}{\cos\theta} \right)\, .
\end{equation}
\noindent
In the case of \J we have estimated $q \simeq 1.078$, leading to $v_{\rm h} \simeq 0.0375 \, c /\cos\theta$ and thus implying $\theta<88\degr$. If we further assume lobes' inclination in the source $\theta\sim 80\degr$, the expansion velocity reads as $v_{\rm h}\sim 0.216\,c$, and hence the source age as $t \simeq D / (2\,v_{\rm h}) \sim 35.4$\,Myr. Thus, the kinematic estimates of the age and of the advance velocity of the observed radio structure in \J are fully consistent with the values resulting from the dynamical modeling presented in \S\,3.3.

We note that the jet inclination $\theta \simeq 80\degr$ is in fact more probable than $\theta \simeq 90\degr$ adopted in the fitting procedure also for the other DDRGs analyzed. However, the resulting model parameters are not sensitive to such relatively minor changes in the assumed value of $\theta$, since a physical size of the lobes increases only by $2\%$ when the source inclination is decreased from $90\degr$ to $80\degr$. Finally, we comment on the fact that even higher advance velocity of the jet head has been estimated for the radio-loud quasar 3C\,273 by \citet{sta04}, namely $v_{\rm h} \gtrsim 0.85\,c$, under the hypothesis that the observed large-scale structure in this source is also a manifestation of the restarted jet activity. The model proposed by \citet{sta04}, which was inspired similarly by somewhat unusual morphological properties of 3C\,273 --- namely by the extremely narrow and only one-sided large-scale lobe in the system --- is in fact analogous to the one discussed here for \J.
  
\subsection{Density of the Ambient Medium}

An upper limit for the thermal gas density within cocoons of FR\,II sources may be estimated from the internal depolarization (lack of any signatures for such) of the observed lobes' radio emission. In particular, \citet{gar91} found that the product of the cold gas number density and the magnetic field intensity within the studied lobes is on average $\langle n_{\rm th} \times B \rangle < 5 \times 10^{-3}$\,$\mu$G\,cm$^{-3}$. This, for the typical range $3$\,$\mu$G\,$<B<30$\,$\mu$G found by means of multiwavelength analysis of the non-thermal lobes' emission \citep[including the radio and X-ray data \emph{for normal-size} FR\,II systems; e.g.,][and references therein]{kat05,cro05}, gives very roughly $\langle n_{\rm th}\rangle < (1-10) \times 10^{-4}$\,cm$^{-3}$, or equivalently $\langle \rho_{\rm th} \rangle \simeq m_{\rm p} \, \langle n_{\rm th}\rangle < (0.3-3) \times 10^{-27}$\,g\,cm$^{-3}$. Below we assume that same limit holds also for the cocoons of GRGs, and the outer lobes of DDRGs.

On the other hand, the upper limit on the plasma density within the extended lobes may be provided by analyzing the total amount of ultrarelativistic (or at least mildly relativistic) plasma deposited by the jets into the expanding cocoons. Neglecting the work done by such expanding cocoons on the ambient medium, the total energy density deposited by a pair of jets with kinetic energy $Q_{\rm j}$ over the source lifetime $t$ into the cavity with volume $V=\pi D^{3}/(16\,R_{\rm T}^{2})$ is simply $u_{\rm rel} \simeq 2 \, Q_{\rm j} \, t / V$. This energy density is shared between the lobes' magnetic field, electron-positron pairs, and also possibly protons, $u_{\rm tot} = u_{\rm B} + u_{\rm e} + u_{\rm p}$. For the purpose of rough estimates, below we assume an energy equipartition between different plasma species within the evolved lobes, as well as mixed (hadronic-leptonic) jet composition, leading to $u_{\rm rel} \simeq 3\, u_{\rm p} \equiv 3\, \langle\gamma_{\rm p}\rangle \, m_{\rm p}c^{2} \, n_{\rm rel}$, where $\langle\gamma_{\rm p}\rangle$ is a mean Lorentz factor of the protons within the lobes, and $n_{\rm rel}$ is their number density. Noting that in the case of an electron-proton jet propagating with relativistic bulk velocities, for which a significant/dominant fraction of the kinetic energy is carried by the particles \citep[as in fact suggested by the most recent analysis for luminous quasar sources; e.g.,][]{sik00,kat08}, protons injected into the lobes through the jet termination shock are expected to be characterized by mean energies $\langle\gamma_{\rm p}\rangle \simeq \Gamma_{\rm j}$, where $\Gamma_{\rm j}$ is the bulk Loretz factor of the outflow \citep[see, e.g., the discussion in][]{kin04,sta07}. Hence the minimum number density of the lobes reads as
\begin{equation}
n_{\rm rel} \simeq \frac{2 \, Q_{\rm j} \, t}{3 \, V \, \Gamma_{\rm j} \, m_{\rm p} c^2} \, .
\end{equation}
\noindent
For the parameters of the hypothetical relict lobes of \J implied by the ``Outer'' model (Table\,6), namely $Q_{\rm j}=1.2\times 10^{46}$\,erg\,s$^{-1}$, $t=178$\,Myr, $D=4.7$\,Mpc and $R_{\rm T} =6$ (giving the total volume of the cocoon $V \simeq 1.7\times 10^{73}$\,cm$^{3}$), assuming in addition conservatively large-scale jet bulk Lorentz factor $\Gamma_{\rm j} \simeq 3$ \citep[see][]{war97}, we obtain $n_{\rm rel} \sim 6 \times 10^{-10}$\,cm$^{-3}$, or equivalently $\rho_{\rm rel} \simeq m_{\rm p} \, n_{\rm rel} \sim 10^{-33}$\,g\,cm$^{-3}$. This value is in a very good agreement with the analogous one estimated by \citet{saf08} for the outer lobes of DDRG PKS\,B1545$-$321 (J1548$-$3216 in this paper).

To sum up, the dynamical modeling presented in \S\,3 implies that the density of the ambient medium around the observed lobes of \J is $\rho_{\textrm{\scriptsize ``inn''}}^{\rm J1420} \sim 2 \times 10^{-31}$\,g\,cm$^{-3}$. The hypothetical outer lobes in the system would be instead surrounded by the gas with the density $\rho_{\textrm{\scriptsize ``out''}}^{\rm J1420} \sim 10^{-29}$\,g\,cm$^{-3}$. These values can be compared with the densities of the environment found in the modeling for of the outer lobes in the control sample of DDRG, $\rho_{\rm out}^{\rm DDRGs} \sim (0.3-3) \times 10^{-28}$\,g\,cm$^{-3}$, and also with the densities of the plasma into which the inner lobes of the studied DDRGs are evolving (i.e., with the densities inside the outer cocoons in the systems), namely $\rho_{\rm inn}^{\rm DDRGs} \sim (0.3-3) \times 10^{-27}$\,g\,cm$^{-3}$. Note that the latter densities are already very close to/almost the same as the upper limits for the thermal gas density $\langle \rho_{\rm th} \rangle$ provided by the analysis of the internal depolarization effects within the FR\,II lobes. The comparison indicates that even though the environment of the presumed relicts in \J was set to resemble the environments of large-scale structures in DDRGs, $\rho_{\textrm{\scriptsize ``out''}}^{\rm J1420} \lesssim \rho_{\rm out}^{\rm DDRGs}$, the medium surrounding the observed lobes in \J seems much underdense with respect to the medium surrounding inner lobes of DDRGs, $\rho_{\textrm{\scriptsize ``inn''}}^{\rm J1420} \ll \rho_{\rm inn}^{\rm DDRGs}$. Different sizes of the structures have to be kept in mind, though. The density $\rho_{\textrm{\scriptsize ``inn''}}^{\rm J1420}$ emerging from the modeling is however still orders of magnitude above the minimum density of the relict cocoon provided by the injection of electron-proton plasma during the hypothetical previous epoch of the jet activity in the source, $\rho_{\rm rel}$. Thus, a substantial entrainment of matter is required for the relict lobes of \J in the framework of the intermitted jet activity scenario \citep[see the related discussion in][]{kai00,saf08}.

It is worth noting that particularly low density of the gas surrounding the observed lobes in \J is consistent with the mean density of the baryonic matter in the Universe. In fact, assuming that half of the baryons in the Universe are located in unvirialized filamentary WHIM \citep{cen99}, this mean density may be estimated as
\begin{equation}
\langle \rho_{\rm b} \rangle \simeq 0.5 \times \Omega_{\rm b} \times \rho_{\rm cr} \sim 2.5 \times 10^{-31} \, {\rm g\,cm^{-3}} \, ,
\end{equation}
\noindent
where $\Omega_{\rm b} \simeq 0.05$ is the baryonic fraction of the flat $\Lambda$CDM Universe, and $\rho_{\rm cr} \simeq 10^{-29}$\,g\,cm$^{-3}$ is the critical density \citep[e.g.,][]{fuk04}. The fact that $\rho_{\textrm{\scriptsize ``inn''}}^{\rm J1420} \sim \langle \rho_{\rm b} \rangle$ implies that if the recurrent jet activity in \J \emph{is not the case}, the source has to be located instead in a real void of the galaxy and matter distribution, at the outskirts of a filamentary WHIM \citep[cf.][]{sub08,saf09}, since WHIM is expected to constitute matter overdensities $10-30$ times above the mean value $\langle \rho_{\rm b} \rangle$.

\subsection{Supersonic Expansion?}

Extended lobes of GRGs, even though being in many cases relict structures, may be still expanding supersonically with respect to the ambient medium, due to their relatively high internal pressures and low-density environments \citep{sub08}. With no good constraints on the temperature of the non-virialized gas far away from groups and cluster of galaxies, it is however hard to evaluate precisely the sound speed of IGM of interest. For the purpose of rough estimates we write
\begin{equation}
c_{\rm s,\,IGM} \simeq \left( \hat{\gamma}_{\rm IGM} \, \frac{p_{\rm IGM}}{\rho_{\rm IGM}} \right)^{1/2} = \left( \frac{5 k\, T_{\rm IGM}}{3 m_{\rm p}} \right)^{1/2} \simeq 10^7 \, T_6^{1/2} \, {\rm cm\,s^{-1}} \,
\end{equation}
\noindent
where $p_{\rm IGM}$, $\rho_{\rm IGM}$ and $T_{\rm IGM} \equiv T_6 \times 10^6$\,K are the pressure, density, and temperature of the gas surrounding the lobes. Meanwhile, the sideway expansion of the lobes, driven by the cocoon's internal pressure, is roughly $v_{\ell} \simeq (p_{\rm c} /\rho_{\rm IGM})^{1/2}$, and the related Mach number of a bow shock driven in the IGM by the expanding cocoon is ${\cal M}_{\ell} = v_{\ell} / c_{\rm s,\,IGM}$. 

If the currently observed lobes of \J are due to a primary jet activity, i.e. if these lobes evolve within `undisturbed' environment, the expected ambient medium (WHIM) temperature is $0.1 \lesssim T_6 \lesssim 1$ \citep{cen99}. This, together with $\rho_{\rm IGM} \sim \langle \rho_{\rm b} \rangle$ and $p_{\rm c} \sim (1-4) \times 10^{-14}$\,dyn\,cm$^{-2}$, as indicated by the modeling discussed in the previous sections, gives us $v_{\ell} \sim 0.01\,c$, and ${\cal M}_{\ell} \sim (20 - 120)$. If the observed lobes are due to the recurrent jet activity instead, the involved shock Mach number would be reduced due to the expected larger temperature of the IGM heated at the bow shock driven by the outer cocoon during the previous epoch of the jet activity. For this outer cocoon, taking again $T_6 \sim (0.1- 1)$, we obtain $v_{\ell} \lesssim 0.01\,c$, and ${\cal M}_{\ell} \sim (20 - 50)$. Hence, regardless on the details of the jet duty cycle in the considered source, we conclude that the giant lobes in the system are expected to modify substantially the surrounding matter by driving strong shocks and heating WHIM at the outskirts of the filamentary galactic distribution.

\subsection{Energetics}

Relativistic jets in luminous radio galaxies and radio-loud quasars are widely believed to be powered via the \citet{bla77} mechanism by a rotational energy of supermassive black holes (SMBHs) residing in the centers of their host galaxies \citep[see, e.g.,][and references therein]{sik07}. The maximum energy that can be extracted in this way is equal to the `reducible mass' of a maximally-spinning SMBH, namely
\begin{equation}
E_{\rm rot} \simeq 0.3 \, M_{\rm BH} c^2 \simeq 5 \times 10^{61} M_8 \, {\rm erg} \, ,
\end{equation}
\noindent
where $M_{\rm BH} \equiv M_8 \times 10^8 M_{\odot}$ is the black hole mass. The total power of a relativistic outflows formed in the Blandford \& Znajek process is typically approximated as
\begin{equation}
Q_{\rm BZ} \simeq \frac{c}{4\pi}\,\left(\frac{J}{J_{\rm max}}\right)^{2} \, B_{\rm g}^{2} \, r_{\rm g}^{2} \, ,
\end{equation}
\noindent
where $J/J_{\rm max}$ is the black hole spin normalized to the maximum value $J_{\rm max} = G M_{\rm BH}^2/c$, the intensity of the magnetic field threading the black hole horizon is $B_{\rm g}$, and $r_{\rm g} = G M_{\rm BH} / c^2$ is the gravitational radius of a SMBH \citep[for the most recent discussion on this issue see][]{tch10}. The exact value of $B_{\rm g}$ is not known, but it is often assumed that the maximum magnetic field intensity close to the black hole horizon corresponds to the case when the magnetic field energy density is equal to the radiation energy density of a system accreting at the Eddington rate. This gives $B_{\rm g} \simeq (2 \, L_{\rm Edd} / r_{\rm g}^2 \, c)^{1/2} \simeq 6 \times 10^4 M_8^{-1/2}$\,G, where $L_{\rm Edd} = 4 \pi G M_{\rm BH} m_{\rm p} c / \sigma_{\rm T} \simeq 10^{46} M_8$\,erg\,s$^{-1}$ is the Eddington luminosity. Combining all the above formulae together, and assuming further $J = J_{\rm max}$, we obtain finally $Q_{\rm BZ} \sim 2 \times 10^{45} \, M_8$\,erg\,s$^{-1}$.

In the case of \J an estimate of the mass of the central SMBH can be derived from the velocity dispersion in optical spectrum of the host galaxy. Such a spectrum, obtained with the ISIS double-beam spectrograph at the WHT 4.2\,m telescope, was published previously in \citet{mach08}. Based on the spectrum, the stellar velocity dispersion is determined using the \St synthesis code of \citet{cid05}. The code fits a linear combination of simple stellar populations taken from the evolutionary models of \citet{bru03} to an observed emission spectrum. Stellar motions are modeled by a Gaussian distribution with the dispersion $v_{\rm d}$, from which one can recover the galaxy stellar velocity dispersion, $\sigma_{*}$. In the fitting procedure we use only the red part of the spectrum, because of a better spectral resolution in the red beam than that in the blue one. The obtained \St fits of the observed absorption lines broadening, when corrected for the instrumental and base resolution, give $\sigma_{*}=194\pm 16$\,km\,s$^{-1}$. A relatively large uncertainty in the above arises from a rather poor signal-to-noise ratio in the available dataset (${\rm S/N} \simeq 9$). Next, using the empirical relation by \citet{tre02}
\begin{equation}
\log \left( \frac{M_{\rm BH}}{M_{\odot}} \right) \simeq 8.13 + 4.02\times \log \left( \frac{\sigma_{*}}{\sigma_{0}} \right) \,
\end{equation}
\noindent
where $\sigma_{0}=200$\,km\,s$^{-1}$, we find the black hole mass in \J as $M_{\rm BH} \simeq 10^{8.1 \pm 0.2} \, M_{\odot}$.

Even though the above black hole mass is subjected to some uncertainty, one can see that for the current jet activity in the source $Q_{\rm j} / Q_{\rm BZ} \gtrsim 1$ and $(Q_{\rm j} \times t)/ E_{\rm rot} \gtrsim 0.04$ within the allowed range of $M_{\rm BH}$. Meanwhile, the presumed previous epoch of the jet activity would be characterized by even more extreme ratios, namely $Q_{\rm j} / Q_{\rm BZ} \gtrsim 3$ and $(Q_{\rm j} \times t)/ E_{\rm rot} \gtrsim 0.7$. Thus, the energetic requirements for the source are severe, almost independently on the particular model considered. Note also that the maximum power of the jet extracted via the \citeauthor{bla77} mechanism as estimated above is just a rough approximation \citep[see the discussion in][]{tch10}, and involves an accretion at the maximum, Eddington rate, needed for supporting the required strong magnetic field near the black hole horizon by the infalling matter. This in turn suggests that the central engine of \J --- with little manifestation of the ongoing accretion-related activity at present --- was once ($30$\,Myr ago?) a luminous AGN, radiating at the rate $L_{\rm AGN} \sim 0.1 \, L_{\rm Edd} \gtrsim 10^{45}$\,erg\,s$^{-1}$. This provides interesting constraints on the AGN duty cycle in early-type galaxies.

\section{Conclusions}

In this paper we consider the possibility that the structure of the largest radio galaxy, \J discovered by \citet{mach08}, is formed by a restarted, rather than a primary jet activity. This hypothesis is motivated by the unusual morphological properties of the source, suggesting almost ballistic propagation of powerful jets in a particularly low-density ambient medium up to $>2$\,Mpc distances from the host galaxy. If the low density of the source environment is due to the previous jet activity indeed, or rather due to a unique location of \J in a void region of the galaxy distribution, is the main question addressed in the analysis. The obtained results and the final conclusions are as follows:
\begin{itemize}
\item New observations of \J conducted with the GMRT and VLA confirm its morphology consisting of only two narrow lobes; no trace of any outer cocoon is therefore detected. We model the newly accessed radio data using the \D algorithm of \citet{mach07} to determine --- for a given set of observables --- the main physical parameters of the jets inflating the lobes and of their environment. Different model fits performed for the observed radio structure imply relatively young age of the source $\sim 35$\,Myr, its relatively high expansion velocity $\sim 0.2\,c$, large kinetic power $\sim 4 \times 10^{45}$\,erg\,s$^{-1}$, and confirm particularly low-density environment $\sim 2 \times 10^{-31}$\,g\,cm$^{-3}$. The kinematic estimates of the age and of the advance velocity are fully consistent with the values resulting from the \D modeling. 
\item We find that it is possible to chose a realistic set of the model parameters for which the hypothetical outer lobes of \J are old enough so that their expected radio surface brightness is substantially below the rms noise level of the available radio maps. We speculate that future high-sensitivity VLA and Low Frequency Array observations, or X-ray studies using facilities such as XMM-{\it Newton} or {\it Suzaku}, may be more promising for investigating a presence of the hypothetical outer cocoon in \J.
\item The extremely low density of the gas surrounding the observed lobes in \J is consistent with the mean density of the baryonic matter in the Universe. This suggests that if the recurrent jet activity in \J is not the case, the source has to be located instead in a real void of the galaxy and matter distribution. This conclusion is partly supported by the galaxy counts performed in the vicinity of the studied object. If correct, it further implies that the giant lobes in \J are expected anyway to modify dramatically their surroundings by driving strong shocks (Mach numbers $>20$) and heating the gas located at the outskirts of the filamentary galactic distribution (`warm-hot phase of the intergalactic medium'). 
\item Finally, we also find that the energetic requirements for the source (with a given mass of the central supermassive black hole estimated in this paper) are severe in terms of the total jet power and the total energy deposited by the outflows far away from the central engine. Rough estimates presented imply therefore that this central engine --- with little manifestation of the ongoing accretion-related activity at present --- was once a luminous AGN, radiating at the maximum (Eddington) rate. This provides interesting constraints on the AGN duty cycle in \emph{isolated} early-type galaxies.
\end{itemize}

\acknowledgements
We thank Dr Chiranjib Konar who performed the GMRT observations of \J and the staff of the GMRT instrument, operated by the National Centre for Radio Astrophysics of the Tata Institute of Fundamental Research. This project was supported in part by the MNiSW funds for scientific research in years 2009--2012 under the contract No. 3812/B/H03/2009/36. The presented work facilitated the MAPS Catalog of POSS-I provided by the University of Minnesota ({\texttt http://aps.umn.edu/}). \L.~S. acknowledges the support from the Polish MNiSW through the grant N-N203-380336.

\clearpage

\begin{table}
\footnotesize
\centering
\caption{Main properties of the studied sources.}
\begin{tabular}{llcrr}
\tableline\tableline
Name		& Redshift 	& 	Type/Lobe    & Total Liner Size    & Lobes' Axial Ratio \\
	 & $z$ & & $D$ [kpc] & $R_{\rm T}$ \\
\tableline
\J	&	0.3067	&	GRG/Northern	&	2257$\pm$9	&	12$\pm$3 \\
	&			&	GRG/Southern	&	2433$\pm$14	&	12$\pm$3 \\
\\
J0041+3224	&	(0.45)	&	DDRG/Outer	&	993$\pm$17	&	4.4$\pm$1.4 \\
			&			&	DDRG/Inner	&	182$\pm$6	&	9.0$\pm$3.0 \\
\\
J1453+3308	&	0.249	&	DDRG/Outer	&	1320$\pm$28	&	3.8$\pm$1.3 \\
			&			&	DDRG/Inner	&	170$\pm$6	&	9.6$\pm$3.2 \\
\\
J1548$-$3216	&	0.1082	&	DDRG/Outer	&	998$\pm$22	&	2.9$\pm$0.5 \\
			&			&	DDRG/Inner	&	317$\pm$4	&	11$\pm$4 \\
\tableline
\end{tabular}
\end{table}

\clearpage

\begin{table}
\footnotesize
\centering
\caption{New GMRT and VLA observations of \J.}
\begin{tabular}{ccrrl}
\tableline\tableline
Telescope 			& Array 		& Frequency       & Primary Beam  & Date \\
       			&       		& $\nu$ [MHz]      & [arc\,min] & \\
\tableline
GMRT                    & F\tablenotemark{a}    & 328.8 & 81        & 2008, May 7  \\
GMRT                    & F     		& 617.3 & 43        & 2008, May 21 \\
VLA\tablenotemark{b}    & D     		& 4860.1     & 9.3    & 2008, Jul 3  \\
VLA\tablenotemark{c}    & D     		& 4860.1     & 9.3    & 2009, Oct 25 \\
\tableline
\end{tabular}
\tablenotetext{a}{``F'' denotes the full array.}
\tablenotetext{b}{Observations of the northern lobe.}
\tablenotetext{c}{Observations of the southern lobe.}
\end{table}

\clearpage

\begin{table}
\footnotesize
\centering
\caption{Instrumental characteristics for the presented radio maps of \J.}
\begin{tabular}{rrrr}
\tableline\tableline
Frequency 			& Beam Size              & Position Angle    & RMS Noise \\
	$\nu$ [MHz] & [arc\,sec] & [deg] & [mJy\,beam$^{-1}$] \\
\tableline
328.8                   &16.1$\times$7.8  &59.8      & 0.103 		  \\
617.3                   &7.1$\times$6.3   &50.0      & 0.035 		  \\
1400.0                  &6.4$\times$5.4   & 0.0      & 0.153              \\
4860.1                  &10.8$\times$17.2 &$-$11.5   & 0.048 		  \\
4860.1                  & 9.9$\times$14.4 &$-$14.8   & 0.036 		  \\
\tableline
\end{tabular}
\end{table}

\clearpage

\begin{table*}
\footnotesize
\centering
\caption{Observed radio fluxes and luminosities of \J including its Northern and Southern lobes, core, as well as of the source confusing the Northern lobe.}
\begin{tabular*}{145mm} {@{}r @{}r @{}r  @{}r  @{}r @{}r @{}r @{}r @{}r @{}r @{}r  @{}r @{}r}
\tableline
\tableline
\multicolumn{2}{c}{} &
\multicolumn{4}{c}{Northern lobe} &
\multicolumn{4}{c}{Southern lobe} &
\multicolumn{2}{c}{Core}	& 
\multicolumn{1}{c}{N-conf.} \\
\tableline
\multicolumn{1}{c}{$\nu$}	&
\multicolumn{1}{c}{} &
\multicolumn{1}{c}{$S_{\nu}$}  &
\multicolumn{1}{c}{} &
\multicolumn{1}{c}{$\log P_{\nu}$} &
\multicolumn{1}{c}{} &
\multicolumn{1}{c}{$S_{\nu}$} &
\multicolumn{1}{c}{} &
\multicolumn{1}{c}{$\log P_{\nu}$} &
\multicolumn{1}{c}{} &
\multicolumn{1}{c}{$S_{\nu}$} &
\multicolumn{1}{c}{} &
\multicolumn{1}{c}{$S_{\nu}$}\\
\multicolumn{1}{c}{[MHz]}	&
\multicolumn{1}{c}{} &
\multicolumn{1}{c}{[mJy]}  &
\multicolumn{1}{c}{} &
\multicolumn{1}{c}{[W\,Hz$^{-1}$\,sr$^{-1}$]} &
\multicolumn{1}{c}{} &
\multicolumn{1}{c}{[mJy]} &
\multicolumn{1}{c}{} &
\multicolumn{1}{c}{[W\,Hz$^{-1}$\,sr$^{-1}$]} &
\multicolumn{1}{c}{} &
\multicolumn{1}{c}{[mJy]} &
\multicolumn{1}{c}{} &
\multicolumn{1}{c}{[mJy]}\\
\tableline
328.8 &  & 131$\pm$18 & & 24.436$\pm$0.060 & & 100$\pm$16 & & 24.318$\pm$0.070 & & $<$1.0 & & 16.0$\pm$2.4 \\
617.3 &  & 86$\pm$8  &  & 24.270$\pm$0.041 & & 67$\pm$7  & & 24.162$\pm$0.046 & & $<$1.3 & & 11.1$\pm$1.8\\
1400.0 &  &  49$\pm$6  & & 24.060$\pm$0.054 & & 41$\pm$5  & & 23.964$\pm$0.053 & & 2.0$\pm$0.2 & & 5.1$\pm$0.7\\
4860.1 &  &  11$\pm$1.2 & & 23.464$\pm$0.048 & &  11.7$\pm$1.8 & & 23.468$\pm$0.068 & & 2.1$\pm$0.3 & & 2.7$\pm$0.4\\
\tableline
\end{tabular*}
\end{table*}

\clearpage

\begin{table*}
\footnotesize
\centering
\caption{Observed radio fluxes and luminosities of the three DDRGs modeled in the paper.}
\begin{tabular*}{137mm} {@{}l @{}r @{}r @{}r  @{}r  @{}r @{}r @{}r @{}r @{}r @{}r}
\tableline
\tableline
\multicolumn{1}{l}{DDRG} &
\multicolumn{1}{c}{} &
\multicolumn{1}{c}{} &
\multicolumn{4}{c}{Outer lobes} &
\multicolumn{4}{c}{Inner lobes} \\
\tableline
\multicolumn{1}{l}{} &
\multicolumn{1}{c}{} &
\multicolumn{1}{c}{$\nu$}	&
\multicolumn{1}{c}{} &
\multicolumn{1}{c}{$S_{\nu}$}  &
\multicolumn{1}{c}{} &
\multicolumn{1}{c}{$\log P_{\nu}$} &
\multicolumn{1}{c}{} &
\multicolumn{1}{c}{$S_{\nu}$} &
\multicolumn{1}{c}{} &
\multicolumn{1}{c}{$\log P_{\nu}$} \\
\multicolumn{1}{l}{} &
\multicolumn{1}{c}{} &
\multicolumn{1}{c}{[MHz]}	&
\multicolumn{1}{c}{} &
\multicolumn{1}{c}{[mJy]}  &
\multicolumn{1}{c}{} &
\multicolumn{1}{c}{[W\,Hz$^{-1}$\,sr$^{-1}$]} &
\multicolumn{1}{c}{} &
\multicolumn{1}{c}{[mJy]} &
\multicolumn{1}{c}{} &
\multicolumn{1}{c}{[W\,Hz$^{-1}$\,sr$^{-1}$]} \\
\tableline
J0041+3224 & &617 &    &   1140$\pm$120$^{a}$ & & 25.826$\pm$0.046 & &  1005$\pm$70$^{a}$ & & 25.733$\pm$0.030\\
& &1287 &    &  460$\pm$35$^{a}$  & & 25.466$\pm$0.033 & &  518$\pm$26$^{a}$ & & 25.445$\pm$0.022\\
& &1400 &    &  409$\pm$43$^{a}$  & & 25.416$\pm$0.046 & &  525$\pm$37$^{a}$ & & 25.451$\pm$0.031\\
& &4860 &    &  87$\pm$7$^{a}$   &  & 24.770$\pm$0.035 & &  206$\pm$10$^{a}$ & & 25.044$\pm$0.022\\
& &8460 &    &  $>$20 &  &  ---  &   &   132$\pm$7$^{a}$  & & 24.851$\pm$0.023\\
\tableline
J1453+3308& & 240 &    &  1657$\pm$249$^{b}$ & & 25.349$\pm$0.066 & & 176$\pm$26$^{c}$ & & 24.375$\pm$0.065\\
& &334 &    &  1465$\pm$154$^{b}$ & & 25.304$\pm$0.046 & & 141$\pm$12$^{c}$ & & 24.280$\pm$0.037\\
& &605 &    &  976$\pm$73$^{b}$  & & 25.144$\pm$0.033 & & 105$\pm$10$^{c}$ & & 24.154$\pm$0.040\\
& &1287 &    &  436$\pm$33$^{b}$  & & 24.811$\pm$0.033 & & 58$\pm$9$^{c}$  & & 23.900$\pm$0.073\\
& &1400 &    &  403$\pm$30$^{b}$  & & 24.778$\pm$0.034 & & 57$\pm$6$^{c}$  & & 23.893$\pm$0.046\\
& &4860 &    &  103$\pm$8$^{b}$   & & 24.203$\pm$0.035 & & 27$\pm$4$^{c}$  & & 23.579$\pm$0.065\\
& &8460 &    &   ---  &    &  ---   &    &  14$\pm$3$^{c}$  & & 23.301$\pm$0.094\\
\tableline
J1548$-$3216 & & 160 &    & 8400$\pm$840$^{d}$ & & 25.279$\pm$0.045 & & --- & & --- \\
& &334 &    & 4737$\pm$710$^{d}$ & & 25.034$\pm$0.066 & & 188$\pm$29$^{d}$ & & 23.632$\pm$0.067\\
& &619 &    & 3141$\pm$252$^{d}$ & & 24.859$\pm$0.035 & & 131$\pm$12$^{d}$ & & 23.475$\pm$0.040\\
& &843 &    & 2410$\pm$180$^{e}$ & & 24.747$\pm$0.033 & & --- & & --- \\
& &1384 &   & 1733$\pm$87$^{d}$  & & 24.608$\pm$0.022 & & 80$\pm$7.3$^{d}$ & & 23.261$\pm$0.040\\
& &2495 &   & 963$\pm$45$^{d}$  & & 24.360$\pm$0.021 & &  53$\pm$6.2$^{d}$ & & 23.082$\pm$0.051\\
& &4860 &   & 415$\pm$42$^{d}$  & & 24.006$\pm$0.044 & &  31.5$\pm$3.8$^{d}$ & & 22.856$\pm$0.053\\
& &4910 &    &  ---   &    &   ---  &     &      35.0$\pm$2.4$^{d}$ & & 22.902$\pm$0.030\\
\tableline
\end{tabular*}
\begin{flushleft}
$^{a}$Flux densities from Table\,2 of \citet{sai06}.\\
$^{b}$Flux densities from Table\,2 of \citet{kon06}.\\
$^{c}$Flux densities from Table\,3 of \citet{kon06} after subtraction of the core emission.\\
$^{d}$Flux densities from Table\,2 of \citet{mach10}.\\
$^{e}$Flux density adopted from the Molonglo Reference Catalogue \citep[MRC;][]{jon92} after subtracting the estimated flux density of the inner lobes made by interpolation between the values in column (6).
\end{flushleft}
\end{table*}

\clearpage

\begin{table*}
\scriptsize
\centering
\caption{Model fit parameters for the analyzed sources.}
\begin{tabular*}{165mm}{llcccccccccccc}
\tableline
\tableline
& &\multicolumn{2}{c}{J0041+3224} & & 
\multicolumn{2}{c}{J1453+3308} & & 
\multicolumn{2}{c}{J1548$-$3216}  & & 
\multicolumn{3}{c}{J1420$-$0545}\\
& & Outer  & Inner  && Outer  &  Inner && Outer & Inner && ``Primary'' & ``Inner'' & ``Outer''\\
\tableline
A) & $a_{0}$\,[kpc]	& 10  & 10  && 10  & ---    &&  10 & ---    && 10  & 10    & 10\\
& $\beta$		& 1.5  & 0.1  && 1.5  & 0    &&  1.5 & 0    && 1.5  & 0.1    & 1.5\\
& $\theta$\,[deg] & 90  & 90  && 90  & 90    &&  90 & 90    && 90  & 90    & 90\\
& $k^{\prime}$		& 10  & 0  && 10  & 0    &&  10 & 0    && 1  & 0    & 10\\
& $\gamma_{\rm min}$	& 1  & 1  && 1  & 1    &&  1 & 1    && 1  & 1    & 1\\
& $\gamma_{\rm max}$	& $10^7$  & $10^7$  && $10^7$  & $10^7$    &&  $10^7$ & $10^7$    && $10^7$  & $10^7$    & $10^7$\\
& $\hat{\gamma}_{c}$, $\hat{\gamma}_{\rm IGM}$ & 5/3  & 5/3  && 5/3  & 5/3    &&  5/3 & 5/3    && 5/3  & 5/3    & 5/3\\
\tableline
B) & $Q_{\rm j}$\,[$10^{45}$\,erg\,s$^{-1}$]		& 14.4 & 2.49 && 3.92 & 0.32 && 1.59 & 0.14 && 4.57 & 3.95  & 12.0$^f$\\
& $\rho_{0}$\,[$10^{-27}$\,g\,cm$^{-3}$] & 10.8 & 2.64 && 73.0 & 0.94 && 78.9 & 0.15 && 0.55 &0.00033& 40.0$^f$\\
& $\alpha_{\rm inj}$                         & 0.620& 0.600&& 0.501& 0.660&& 0.505& 0.608&& 0.547& 0.534 & 0.54\\
& $t$\,[Myr]                                   & 105  & 4.0  && 104  & 5.0  && 132  & 9.2  &&  35  &  32    & 178\\
& $t_{\rm j}$\,[Myr]                           &  94  & ---  &&  80  & ---  && 102  & ---  && ---  &  ---   & 106\\
& $\chi^{2}_{\rm red.}$                      & 0.74 & 2.20 && 0.98 & 0.35 && 1.04 & 0.60 && 0.95 & 0.85  & ---\\
\tableline
C) & $v_{\rm h}(t)/c$                                   &0.013 &0.046 &&0.018 &0.030 &&0.011 &0.031 && 0.186& 0.146  & 0.037\\
& $D/(2c\,t)$                                  &0.015 &0.075 &&0.021 &0.055 &&0.012 &0.056 && 0.218& 0.239  & 0.043\\
& $\rho(D)$\,[$10^{-29}$\,g\,cm$^{-3}$] & 3.09 & ---  && 13.6 & ---  && 22.4 & ---  &&0.015 & 0.019 & 1.11\\
& $p_{\rm c}$\,[$10^{-13}$\,dyn\,cm$^{-2}$] & 227  & 170  && 23.0 & 32.0 && 21.1 & 4.00 && 0.42 & 0.13   & 3.80\\
& $B$\,[$\mu$G]                            & 7.7 & 24.0 && 2.3 & 15.5 && 2.2 & 3.7 && 0.51& 0.65  & 0.97\\
\hline
\end{tabular*}
\begin{flushleft}
A) Parameters $a_0$, $\beta$, $\theta$, $k^{\prime}$, $\gamma_{\rm min}$, $\gamma_{\rm max}$, $\Gamma_{c}$ and $\Gamma_{\rm igm}$ are fixed during the fitting procedure to the values given in the table.\\
B) Parameters $Q_{\rm j}$, $\rho_{0}$, $\alpha_{\rm inj}$, $t$, and $t_{\rm j}$ are the results of the model fitting (with $\chi^{2}_{\rm red.}$ given in the table), except of the cases marked by a superscript $f$ indicating the values assumed (see the discussion in \S\,3.3).\\
C) Quantities $v_{\rm h}(t)/c$, $D/(2c\,t)$, $\rho(D)$, $p_{\rm c}$, and $B_{\rm c}$ are evaluated based on the values of the model parameters given above in the table.\\
$^f$Parameters fixed in the fitting procedure (see the discussion in \S\,3.3).
\end{flushleft}
\end{table*}

\clearpage

\begin{figure}
\begin{center}
\includegraphics[angle=0,scale=1.4]{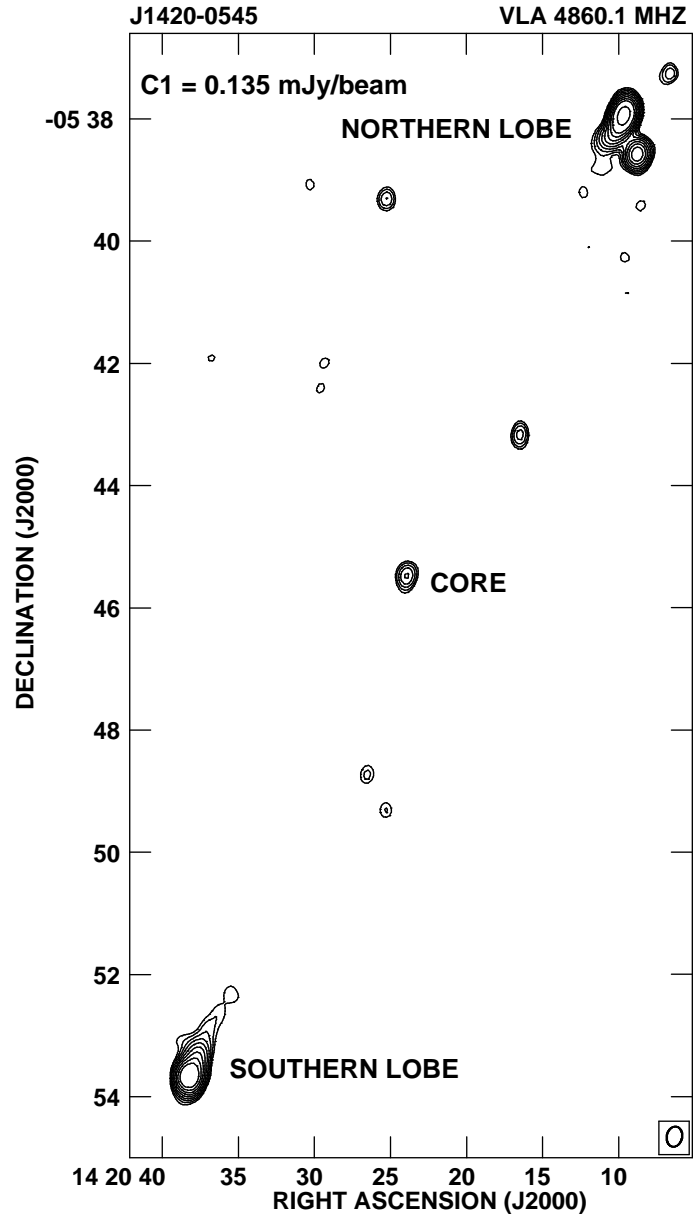}
\end{center}
\caption{VLA image of \J at $4860.1$\,MHz with an angular resolution convolved to $15\arcsec\times20\arcsec$, shown as an ellipse in the bottom right-hand corner. The position of the radio core, as well as the northern and southern lobes, are also marked. The contours spaced by a factor of $\sqrt2$ in brightness are plotted starting with a value of 0.135\,mJy\,beam$^{-1}$.\label{fig1}}
\end{figure}

\clearpage

\begin{figure}
\begin{center}
\includegraphics[angle=0,scale=0.8]{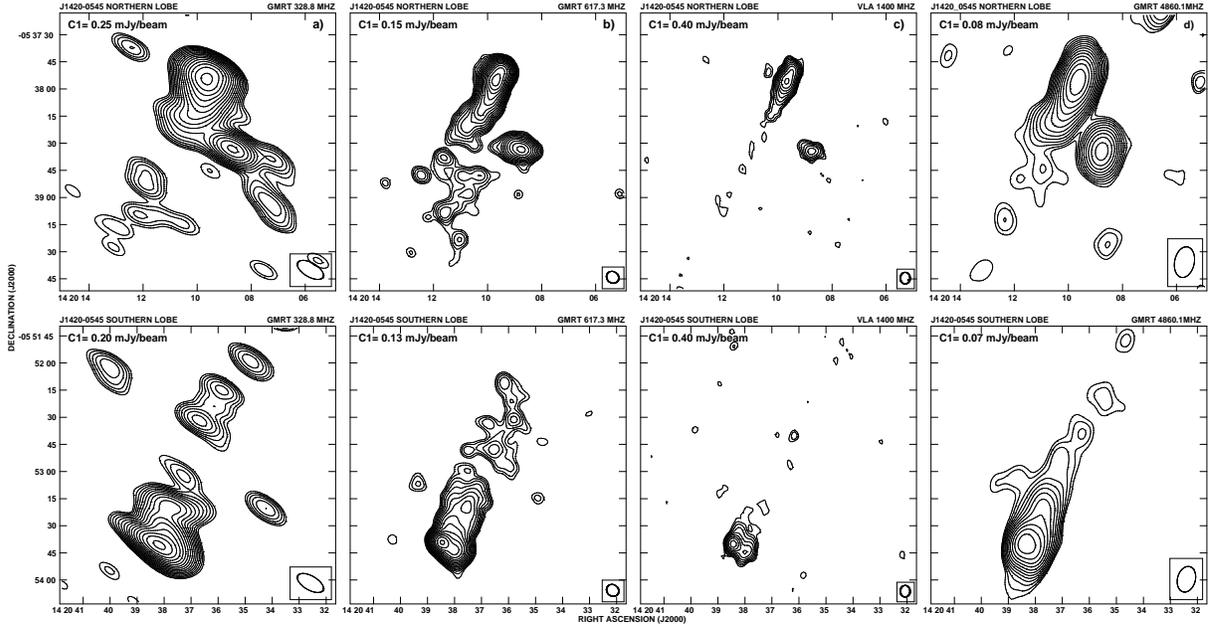}
\end{center}
\caption{Northern and southern lobes of \J (top and bottom panels, respectively), as observed at four different frequencies: (a) $328.8$\,MHz with GMRT, (b) $617.3$\,MHz with GMRT, (c) $1400$\,MHz with VLA (FIRST survey), and (d) $4860.1$\,MHz with VLA. The contours spaced by factors of $\sqrt2$ in brightness are plotted starting with a value C1 given in each panel in the units of mJy\,beam$^{-1}$. The sizes of the beams are indicated by ellipses in the bottom-right corners of each image.\label{fig2}}
\end{figure}

\clearpage

\begin{figure}
\begin{center}
\includegraphics[angle=0,scale=0.95]{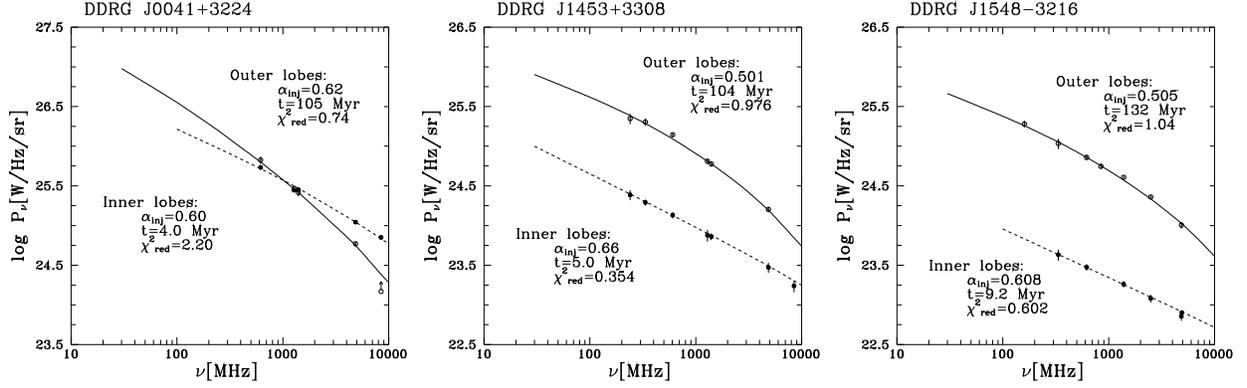}
\end{center}
\caption{Model fits to the comoving spectra of the outer and inner lobes of the three analyzed DDRGs.}
\end{figure}

\clearpage

\begin{figure}
\begin{center}
\includegraphics[angle=0,scale=1.6]{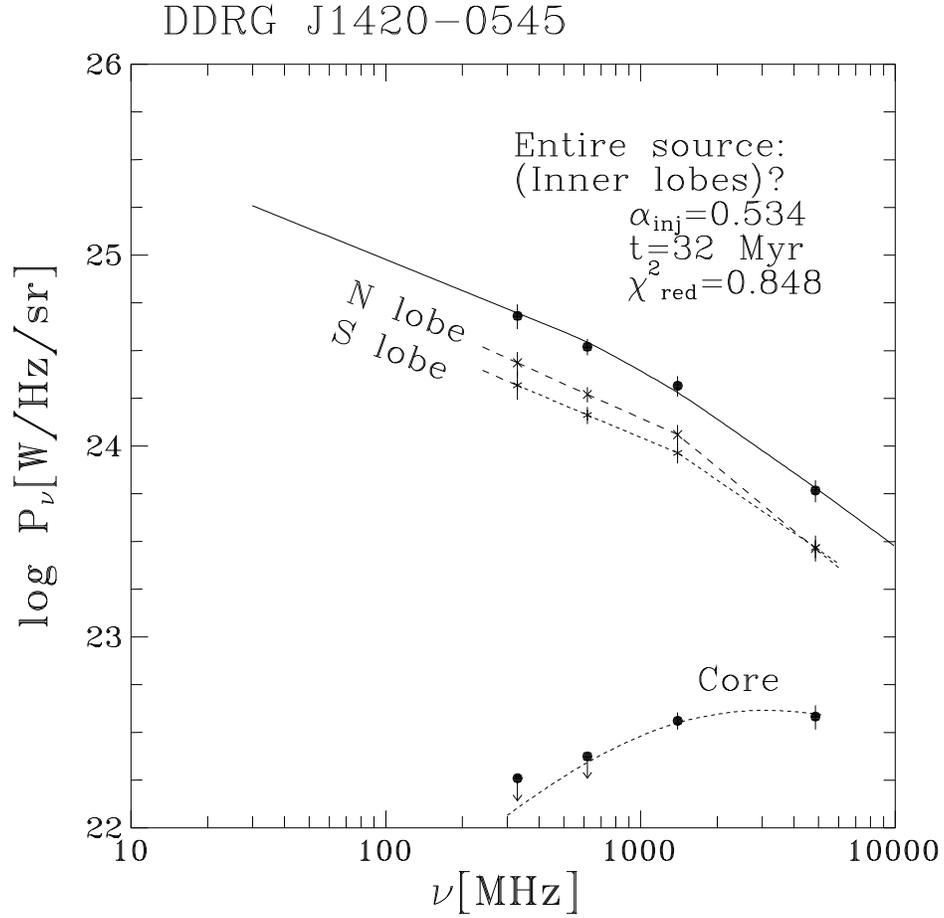}
\caption{``Inner'' model fit to the comoving spectrum of the entire \J structure (solid curve). The spectra of both lobes and of the core are also shown (dashed and dotted curves).}
\end{center}
\end{figure} 

\clearpage

\begin{figure}
\begin{center}
\includegraphics[angle=0,scale=0.8]{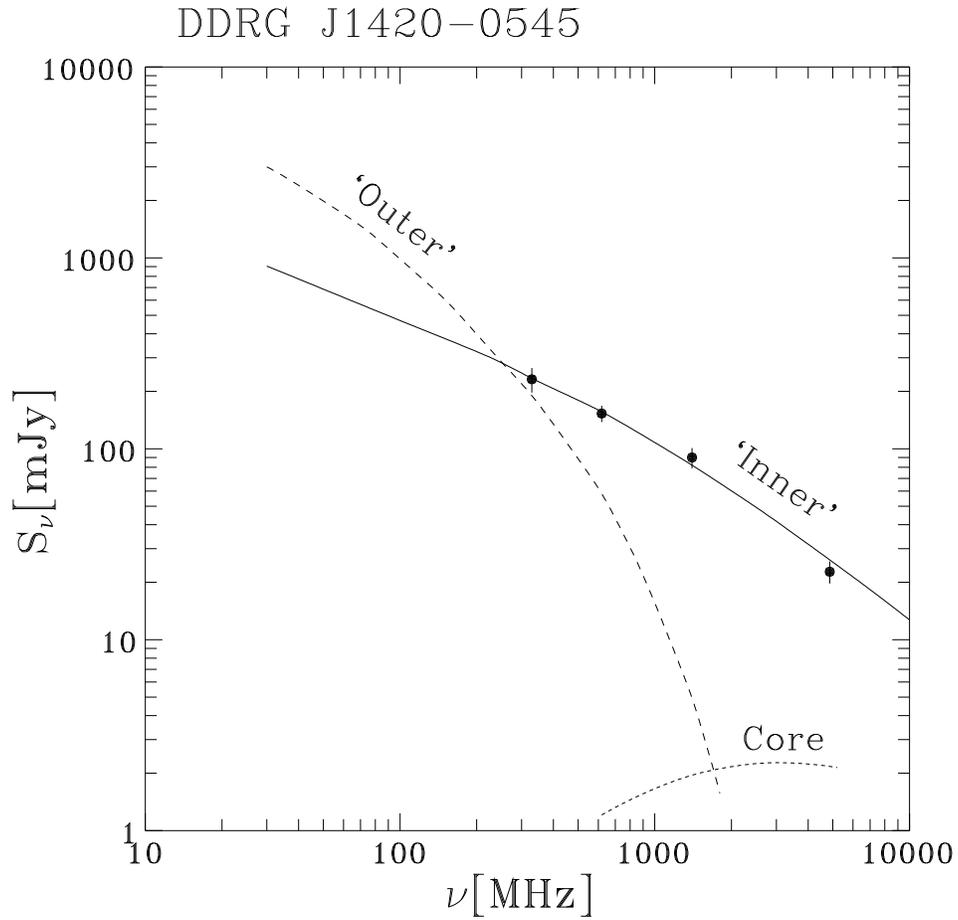}
\caption{``Inner'' model fit to the observed spectrum of the entire \J structure (solid curve), and the expected spectrum of the hypothetical relict lobes of the source corresponding to the ``Outer'' model parameters (dashed curve). The spectrum of radio core is depicted by a dotted curve.}
\end{center}
\end{figure} 

\clearpage

\begin{figure}
\begin{center}
\includegraphics[angle=0,scale=0.95]{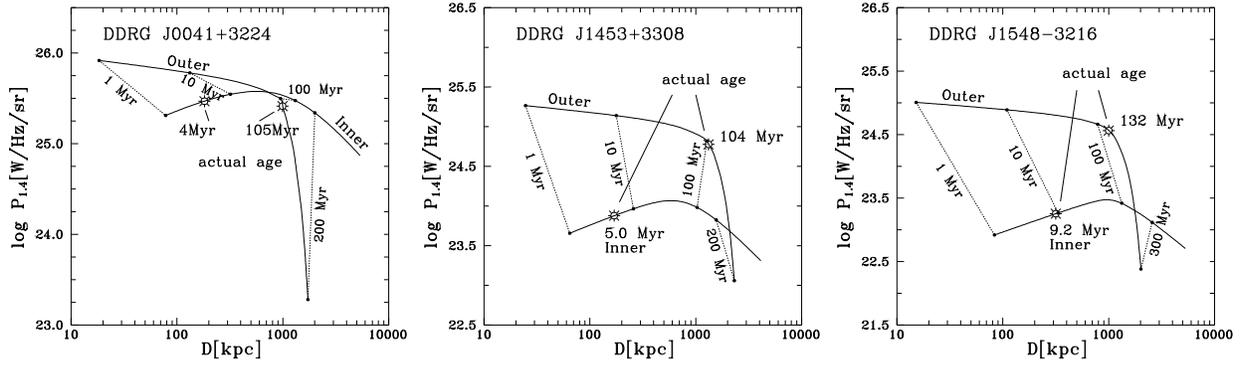}
\caption{Time evolution of the outer and inner lobes of the analyzed DDRGs depicted on the luminosity--size diagrams $P_{\rm 1.4\,GHz} - D$. The actual ages of the radio structures emerging from the model fits are marked by the star symbols, while the consecutive age markers are connected with dotted lines.}
\end{center}
\end{figure}

\clearpage

\begin{figure}
\begin{center}
\includegraphics[angle=0,scale=0.8]{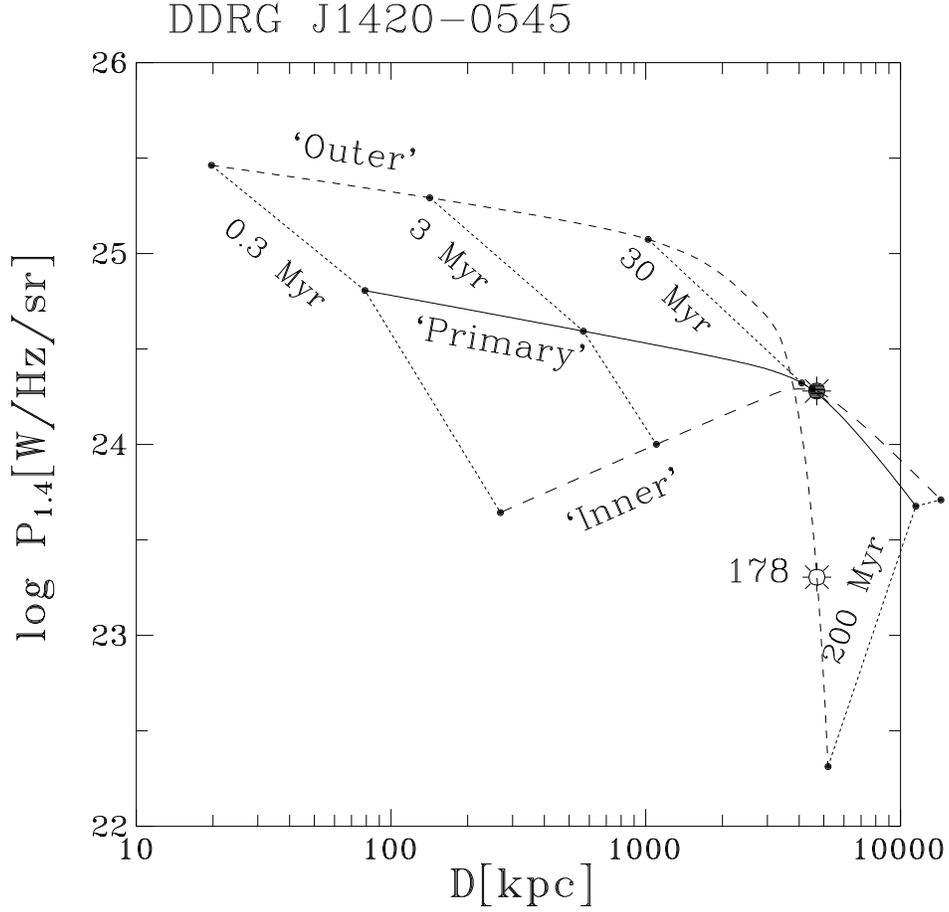}
\caption{Time evolution of \J depicted on the luminosity--size diagrams $P_{\rm 1.4\,GHz} - D$. The tracks corresponding to the ``Primary'' and ``Inner'' models for the observed radio structure are denoted by solid and dashed curves, respectively. The evolutionary track for the hypothetical outer lobes in \J (corresponding to the ``Outer'' model parameters) is given by short-dashed curve. The actual ages of the observed and of the hypothetical/outer radio structures emerging from the model fits are marked by the filled and open star symbols, respectively, while the consecutive age markers are connected with dotted lines.}
\end{center}
\end{figure} 

\clearpage

\begin{figure}
\centering
\includegraphics[angle=0, scale=0.8]{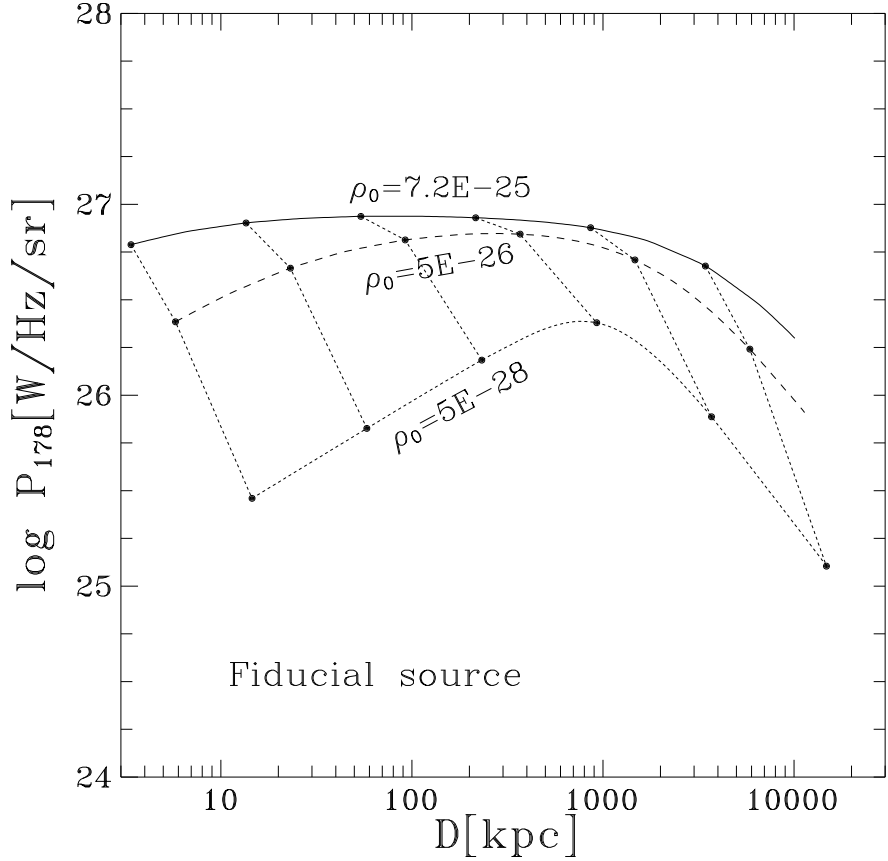}
\caption{$P_{\rm 178\,MHz} - D$ evolutionary tracks of a fiducial source \citep[as considered by][]{KDA97} evolving within the uniform ambient medium ($\beta$=0) characterized by three different values of the density $\rho_{0} = 7.2 \times 10^{-25}$\,g\,cm$^{-3}$ (solid line), $5 \times 10^{-26}$\,g\,cm$^{-3}$ (dashed line), and $5 \times 10^{-28}$\,g\,cm$^{-3}$ (dotted line). The consecutive age markers (dots), connected with dotted lines, correspond to the source ages from 0.1\,Myr to 10\,Gyr.}
\end{figure}

\clearpage

\begin{figure}[]
\centering
\includegraphics[angle=0, scale=0.8]{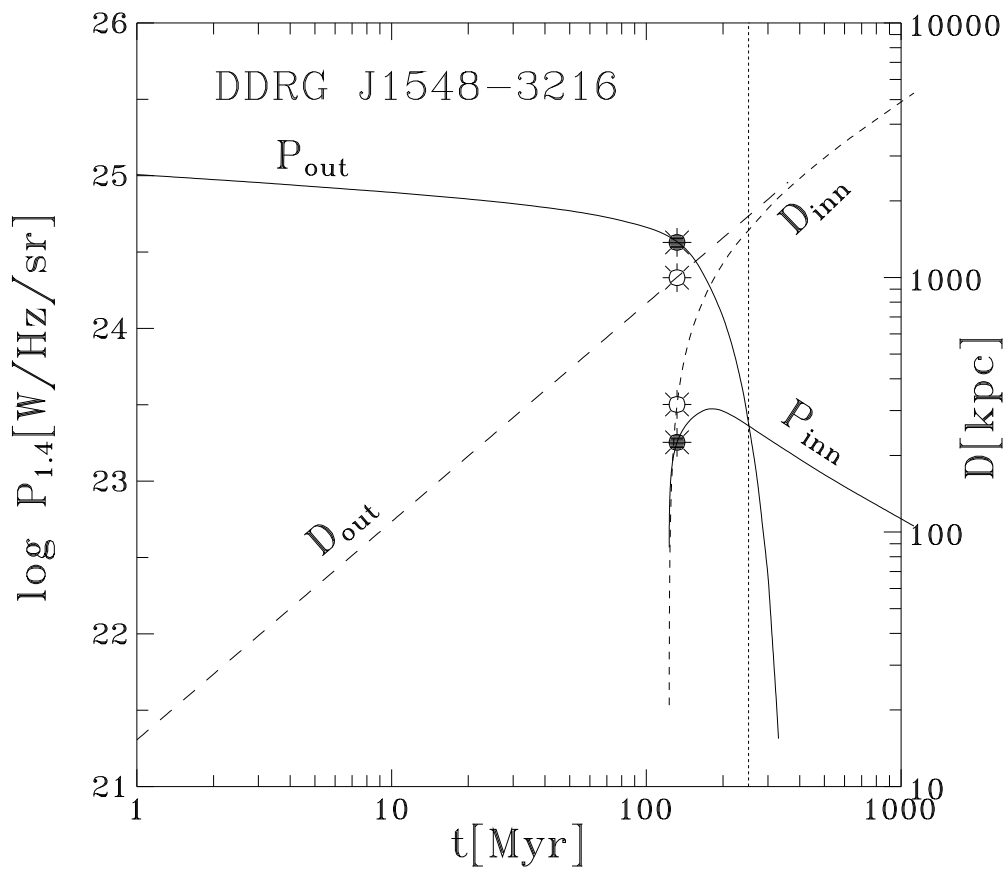}
\caption{Time dependence of a linear size (dashed lines) and luminosity (solid lines) for the outer and inner lobes of DDRG J1548$-$3216. The currently observed sizes and 1400\,MHz luminosities of the structures implied by the model fits are indicated by open and filled star symbols, respectively. The vertical line marks the particular age of the outer lobes starting from which their radio luminosity falls below the radio luminosity of the inner lobes.}
\end{figure}


\begin{thebibliography}{}

\bibitem[Alexander(2000)]{alex00} Alexander, P.\ 2000, \mnras, 319, 8 

\bibitem[Baars et al.(1977)]{baa77} Baars, J.~W.~M., Genzel, R., Pauliny-Toth, I.~I.~K., \& Witzel, A.\ 1977, \aap, 61, 99 

\bibitem[Baldwin(1982)]{bal82} Baldwin, J.~E.\ 1982, Extragalactic Radio Sources, 97, 21 

\bibitem[Becker et al.(1995)]{bec95} Becker, R.~H., White, R.~L., \& Helfand, D.~J.\ 1995, \apj, 450, 559 

\bibitem[Begelman \& Cioffi(1989)]{beg89} Begelman, M.~C., \& Cioffi, D.~F.\ 1989, \apjl, 345, L21 

\bibitem[Best et al.(1995)]{bes95} Best, P.~N., Bailer, D.~M., Longair, M.~S., \& Riley, J.~M.\ 1995, \mnras, 275, 1171 

\bibitem[Blandford \& Rees(1974)]{bla74} Blandford, R.~D., \& Rees, M.~J.\ 1974, \mnras, 169, 395 

\bibitem[Blandford \& Znajek(1977)]{bla77} Blandford, R.~D., \& Znajek, R.~L.\ 1977, \mnras, 179, 433 

\bibitem[Blundell et al.(1999)]{blu99} Blundell, K.~M., Rawlings, S., \& Willott, C.~J.\ 1999, \aj, 117, 677 

\bibitem[Blundell \& Fabian(2011)]{blu11} Blundell, K.~M., \& Fabian, A.~C.\ 2011, \mnras, 412, 705 

\bibitem[Brocksopp et al.(2007)]{bro07} Brocksopp, C., Kaiser, C.~R., Schoenmakers, A.~P., \& de Bruyn, A.~G.\ 2007, \mnras, 382, 1019 

\bibitem[Brocksopp et al.(2011)]{bro11} Brocksopp, C., Kaiser, C.~R., Schoenmakers, A.~P., \& de Bruyn, A.~G.\ 2011, \mnras, 410, 484 

\bibitem[Bruzual \& Charlot(2003)]{bru03} Bruzual, G., \& Charlot, S.\ 2003, \mnras, 344, 1000 

\bibitem[Cen \& Ostriker(1999)]{cen99} Cen, R., \& Ostriker, J.~P.\ 1999, \apj, 514, 1 

\bibitem[Cid Fernandes et al.(2005)]{cid05} Cid Fernandes, R., Mateus, A., Sodr{\'e}, L., Stasi{\'n}ska, G., \& Gomes, J.~M.\ 2005, \mnras, 358, 363 

\bibitem[Clarke(1997)]{cla97} Clarke, D.~A.\ 1997, Computational Astrophysics; 12th Kingston Meeting on Theoretical Astrophysics, 123, 255 

\bibitem[Clarke \& Burns(1991)]{cla91} Clarke, D.~A., \& Burns, J.~O.\ 1991, \apj, 369, 308 

\bibitem[Cohen et al.(2007)]{coh07} Cohen, A.~S., Lane, W.~M., Cotton, W.~D., Kassim, N.~E., Lazio, T.~J.~W., Perley, R.~A., Condon, J.~J., \& Erickson, W.~C.\ 2007, \aj, 134, 1245 

\bibitem[Condon et al.(1998)]{con98} Condon, J.~J., Cotton, W.~D., Greisen, E.~W., Yin, Q.~F., Perley, R.~A., Taylor, G.~B., \& Broderick, J.~J.\ 1998, \aj, 115, 1693 

\bibitem[Croston et al.(2005)]{cro05} Croston, J.~H., Hardcastle, M.~J., Harris, D.~E., Belsole, E., Birkinshaw, M., \& Worrall, D.~M.\ 2005, \apj, 626, 733 

\bibitem[Czerny et al.(2009)]{cze09} Czerny, B., Siemiginowska, A., Janiuk, A., Nikiel-Wroczy{\'n}ski, B., \& Stawarz, {\L}.\ 2009, \apj, 698, 840 

\bibitem[Fabian et al.(2009)]{fab09} Fabian, A.~C., Chapman, S., Casey, C.~M., Bauer, F., \& Blundell, K.~M.\ 2009, \mnras, 395, L67 

\bibitem[Falle(1991)]{fal91} Falle, S.~A.~E.~G.\ 1991, \mnras, 250, 581 

\bibitem[Fanaroff \& Riley(1974)]{fan74} Fanaroff, B.~L., \& Riley, J.~M.\ 1974, \mnras, 167, 31P 

\bibitem[Fukugita \& Peebles(2004)]{fuk04} Fukugita, M., \& Peebles, P.~J.~E.\ 2004, \apj, 616, 643 

\bibitem[Garc{\'{\i}}a-Burillo et al.(2007)]{gar07} Garc{\'{\i}}a-Burillo, S., Combes, F., Neri, R., Fuente, A., Usero, A., Leon, S., \& Lim, J.\ 2007, \aap, 468, L71 

\bibitem[Garrington \& Conway(1991)]{gar91} Garrington, S.~T., \& Conway, R.~G.\ 1991, \mnras, 250, 198 

\bibitem[Hill \& Lilly(1991)]{hil91} Hill, G.~J., \& Lilly, S.~J.\ 1991, \apj, 367, 1 

\bibitem[Isobe et al.(2009)]{iso09} Isobe, N., Tashiro, M.~S., Gandhi, P., Hayato, A., Nagai, H., Hada, K., Seta, H., \& Matsuta, K.\ 2009, \apj, 706, 454 

\bibitem[Isobe et al.(2011a)]{iso11a} Isobe, N., Seta, H., Gandhi, P., \& Tashiro, M.~S.\ 2011a, \apj, 727, 82 

\bibitem[Isobe et al.(2011b)]{iso11b} Isobe, N., Seta, H., \& Tashiro, M.~S.\ 2011b, arXiv:1105.3473 

\bibitem[Jamrozy et al.(2004)]{jam04} Jamrozy, M., Klein, U., Machalski, J., \& Mack, K.-H.\ 2004, Multiwavelength AGN Surveys, 431 

\bibitem[Jamrozy et al.(2008)]{jam08} Jamrozy, M., Konar, C., Machalski, J., \& Saikia, D.~J.\ 2008, \mnras, 385, 1286 

\bibitem[Jones \& McAdam(1992)]{jon92} Jones, P.~A., \& McAdam, W.~B.\ 1992, \apjs, 80, 137 

\bibitem[Kaiser(2000)]{k00} Kaiser, C.~R.\ 2000, \aap, 362, 447 

\bibitem[Kaiser \& Alexander(1997)]{kai97} Kaiser, C.~R., \& Alexander, P.\ 1997, \mnras, 286, 215 

\bibitem[Kaiser \& Alexander(1999)]{kai99} Kaiser, C.~R., \& Alexander, P.\ 1999, \mnras, 302, 515 

\bibitem[Kaiser \& Cotter(2002)]{kai02} Kaiser, C.~R., \& Cotter, G.\ 2002, \mnras, 336, 649 

\bibitem[Kaiser \& Best(2007)]{kai07} Kaiser, C.~R., \& Best, P.~N.\ 2007, \mnras, 381, 1548 

\bibitem[Kaiser et al.(1997)]{KDA97} Kaiser, C.~R., Dennett-Thorpe, J., \& Alexander, P.\ 1997, \mnras, 292, 723 

\bibitem[Kaiser et al.(2000)]{kai00} Kaiser, C.~R., Schoenmakers, A.~P., R\"ottgering, H.~J.~A.\ 2000, \mnras, 315, 381 

\bibitem[Kataoka \& Stawarz(2005)]{kat05} Kataoka, J., \& Stawarz, {\L}.\ 2005, \apj, 622, 797 

\bibitem[Kataoka et al.(2008)]{kat08} Kataoka, J., et al.\ 2008, \apj, 672, 787 

\bibitem[Kino \& Takahara(2004)]{kin04} Kino, M., \& Takahara, F.\ 2004, \mnras, 349, 336 

\bibitem[Konar et al.(2004)]{kon04} Konar, C., Saikia, D.~J., Ishwara-Chandra, C.~H., \& Kulkarni, V.~K.\ 2004, \mnras, 355, 845 

\bibitem[Konar et al.(2006)]{kon06} Konar, C., Saikia, D.~J., Jamrozy, M., \& Machalski, J.\ 2006, \mnras, 372, 693 

\bibitem[Konar et al.(2008)]{kon08} Konar, C., Jamrozy, M., Saikia, D.~J., \& Machalski, J.\ 2008, \mnras, 383, 525 

\bibitem[Konar et al.(2009)]{kon09} Konar, C., Hardcastle, M.~J., Croston, J.~H., \& Saikia, D.~J.\ 2009, \mnras, 400, 480 

\bibitem[Lara et al.(1999)]{lar99} Lara, L., M{\'a}rquez, I., Cotton, W.~D., Feretti, L., Giovannini, G., Marcaide, J.~M., \& Venturi, T.\ 1999, \aap, 348, 699 

\bibitem[Longair \& Riley(1979)]{lon79} Longair, M.~S., \& Riley, J.~M.\ 1979, \mnras, 188, 625 

\bibitem[Machalski(2011)]{mach11} Machalski, J.\ 2011, \mnras, 413, 2429 

\bibitem[Machalski \& Jamrozy(2006)]{mach06} Machalski, J., \& Jamrozy, M.\ 2006, \aap, 454, 95 

\bibitem[Machalski et al.(2007)]{mach07} Machalski, J., Chy{\.z}y, K.~T., Stawarz, {\L}., \& Kozie{\l}, D.\ 2007, \aap, 462, 43 

\bibitem[Machalski et al.(2008)]{mach08} Machalski, J., Kozie{\l}-Wierzbowska, D., Jamrozy, M., \& Saikia, D.~J.\ 2008, \apj, 679, 149

\bibitem[Machalski et al.(2009)]{mach09} Machalski, J., Jamrozy, M., \& Saikia, D.~J.\ 2009, \mnras, 395, 812 

\bibitem[Machalski et al.(2010)]{mach10} Machalski, J., Jamrozy, M., \& Konar, C.\ 2010, \aap, 510, 84 

\bibitem[Mack et al.(1998)]{mack98} Mack, K.-H., Klein, U., O'Dea, C.~P., Willis, A.~G., \& Saripalli, L.\ 1998, \aap, 329, 431 

\bibitem[Manolakou \& Kirk(2002)]{man02} Manolakou, K., \& Kirk, J.~G.\ 2002, \aap, 391, 127 

\bibitem[Mocz et al.(2011)]{moc11} Mocz, P., Fabian, A.~C., \& Blundell, K.~M.\ 2011, \mnras, 413, 1107 

\bibitem[O'Dea(1998)]{odea98} O'Dea, C.~P.\ 1998, \pasp, 110, 493 

\bibitem[Reynolds \& Begelman(1997)]{rey97} Reynolds, C.~S., \& Begelman, M.~C.\ 1997, \apjl, 487, L135 

\bibitem[Ry{\'s}(2000)]{rys00} Ry{\'s}, S.\ 2000, \aap, 355, 79 

\bibitem[Safouris et al.(2008)]{saf08} Safouris, V., Subrahmanyan, R., Bicknell, G.~V., \& Saripalli, L.\ 2008, \mnras, 385, 2117 

\bibitem[Safouris et al.(2009)]{saf09} Safouris, V., Subrahmanyan, R., Bicknell, G.~V., \& Saripalli, L.\ 2009, \mnras, 393, 2 

\bibitem[Saikia \& Jamrozy(2009)]{sai09} Saikia, D.~J., \& Jamrozy, M.\ 2009, Bulletin of the Astronomical Society of India, 37, 63 

\bibitem[Saikia et al.(2006)]{sai06} Saikia, D.~J., Konar, C., \& Kulkarni, V.~K.\ 2006, \mnras, 366, 1391 

\bibitem[Saikia et al.(2007)]{sai07} Saikia, D.~J., Konar, C., Jamrozy, M., Machalski, J., Gupta, N., Stawarz, L., Mack, K.-H., \& Siemiginowska, A.\ 2007, The Central Engine of Active Galactic Nuclei, 373, 217 

\bibitem[Saripalli et al.(2002)]{sar02} Saripalli, L., Subrahmanyan, R., \& Udaya Shankar, N.\ 2002, \apj, 565, 256 

\bibitem[Saripalli et al.(2003)]{sar03} Saripalli, L., Subrahmanyan, R., \& Udaya Shankar, N.\ 2003, \apj, 590, 181 

\bibitem[Saripalli et al.(2005)]{sar05} Saripalli, L., Hunstead, R.~W., Subrahmanyan, R., \& Boyce, E.\ 2005, \aj, 130, 896 

\bibitem[Scheuer(1974)]{sch74} Scheuer, P.~A.~G.\ 1974, \mnras, 166, 513 

\bibitem[Scheuer(1995)]{sch95} Scheuer, P.~A.~G.\ 1995, \mnras, 277, 331 

\bibitem[Schoenmakers et al.(1999)]{sch99} Schoenmakers, A.~P., de Bruyn, A.~G., R{\"o}ttgering, H.~J.~A., \& van der Laan, H.\ 1999, \aap, 341, 44 

\bibitem[Schoenmakers et al.(2000a)]{sch00a} Schoenmakers, A.~P., de Bruyn, A.~G., R{\"o}ttgering, H.~J.~A., van der Laan, H., \& Kaiser, C.~R.\ 2000a, \mnras, 315, 371 

\bibitem[Schoenmakers et al.(2000b)]{sch00b} Schoenmakers, A.~P., Mack, K.-H., de Bruyn, A.~G., R{\"o}ttgering, H.~J.~A., Klein, U., \& van der Laan, H.\ 2000b, \aaps, 146, 293 

\bibitem[Schoenmakers et al.(2001)]{sch01} Schoenmakers, A.~P., de Bruyn, A.~G., R{\"o}ttgering, H.~J.~A., \& van der Laan, H.\ 2001, \aap, 374, 861 

\bibitem[Sikora \& Madejski(2000)]{sik00} Sikora, M., \& Madejski, G.\ 2000, \apj, 534, 109 

\bibitem[Sikora et al.(2007)]{sik07} Sikora, M., Stawarz, {\L}., \& Lasota, J.-P.\ 2007, \apj, 658, 815 

\bibitem[Stawarz(2004)]{sta04} Stawarz, {\L}.\ 2004, \apj, 613, 119 

\bibitem[Stawarz et al.(2007)]{sta07} Stawarz, {\L}., Cheung, C.~C., Harris, D.~E., \& Ostrowski, M.\ 2007, \apj, 662, 213 

\bibitem[Stawarz et al.(2008)]{sta08} Stawarz, {\L}., Ostorero, L., Begelman, M.~C., Moderski, R., Kataoka, J., \& Wagner, S.\ 2008, \apj, 680, 911

\bibitem[Subrahmanyan et al.(2008)]{sub08} Subrahmanyan, R., Saripalli, L., Safouris, V., \& Hunstead, R.~W.\ 2008, \apj, 677, 63 

\bibitem[Tchekhovskoy et al.(2010)]{tch10} Tchekhovskoy, A., Narayan, R., \& McKinney, J.~C.\ 2010, \apj, 711, 50 

\bibitem[Tremaine et al.(2002)]{tre02} Tremaine, S., et al.\ 2002, \apj, 574, 740 

\bibitem[Tyson(1988)]{tys88} Tyson, J.~A.\ 1988, \aj, 96, 1 

\bibitem[Wardle \& Aaron(1997)]{war97} Wardle, J.~F.~C., \& Aaron, S.~E.\ 1997, \mnras, 286, 425 

\end{thebibliography}
\end{document}